\documentclass[10pt,a4paper]{article}

\usepackage{times}
\usepackage[centertags]{amsmath}
\usepackage{amssymb,amsfonts,theorem,stmaryrd}
\usepackage{slashbox}
\usepackage{pstricks}
\usepackage{graphicx}
\usepackage[ps,arc,frame]{xy}
\usepackage{psfrag}
\usepackage{slashed}

\allowdisplaybreaks

\theorembodyfont{\slshape}

\theorembodyfont{\rmfamily}

\newcommand{\R}{\mathbb{R}}

\newcommand{\C}{\mathbb{C}}
\newcommand{\Z}{\mathbb{Z}}

\newcommand{\Y}{\mathcal{Y}}
\newcommand{\G}{\mathcal{G}}
\newcommand{\M}{\mathcal{M}}
\newcommand{\HH}{{\mathcal{H}}}

\newcommand{\id}{\operatorname{id}}
\newcommand{\tr}{\operatorname{tr}}

\newcommand{\rk}{\operatorname{rk}}

\newcommand{\vol}{\operatorname{vol}}

\newcommand{\diag}{\operatorname{diag}}

\newcommand{\pp}[1]{\begin{pmatrix} #1 \end{pmatrix}}

\newcommand{\bb}{\begin{eqnarray}}
\newcommand{\ee}{\end{eqnarray}}
\newcommand{\eee}{\nonumber\end{eqnarray}}

\renewcommand{\Box}{$\boxbox$}

\newcommand{\rxyacht}[2]{{\begin{xy} 0;<2mm,0mm>:<0mm,2mm>::0;0,
,(5,-2)*{a}
,(10,-1.8)*{b}
,(15,-2)*{c}
,(20,-2)*{d}
,(25,-2)*{e}
,(30,-2)*{f}
,(35,-2)*{g}
,(40,-2)*{h}
,(2,-5)*{a}
,(2,-10)*{b}
,(2,-15)*{c}
,(2,-20)*{d}
,(2,-25)*{e}
,(2,-30)*{f}
,(2,-35)*{g}
,(2,-40)*{h}
,(5,-5)*\cir(#1,0){}
,(10,-5)*\cir(#1,0){}
,(15,-5)*\cir(#1,0){}
,(20,-5)*\cir(#1,0){}
,(25,-5)*\cir(#1,0){}
,(30,-5)*\cir(#1,0){}
,(35,-5)*\cir(#1,0){}
,(40,-5)*\cir(#1,0){}
,(5,-10)*\cir(#1,0){}
,(10,-10)*\cir(#1,0){}
,(15,-10)*\cir(#1,0){}
,(20,-10)*\cir(#1,0){}
,(25,-10)*\cir(#1,0){}
,(30,-10)*\cir(#1,0){}
,(35,-10)*\cir(#1,0){}
,(40,-10)*\cir(#1,0){}
,(5,-15)*\cir(#1,0){}
,(10,-15)*\cir(#1,0){}
,(15,-15)*\cir(#1,0){}
,(20,-15)*\cir(#1,0){}
,(25,-15)*\cir(#1,0){}
,(30,-15)*\cir(#1,0){}
,(35,-15)*\cir(#1,0){}
,(40,-15)*\cir(#1,0){}
,(5,-20)*\cir(#1,0){}
,(10,-20)*\cir(#1,0){}
,(15,-20)*\cir(#1,0){}
,(20,-20)*\cir(#1,0){}
,(25,-20)*\cir(#1,0){}
,(30,-20)*\cir(#1,0){}
,(35,-20)*\cir(#1,0){}
,(40,-20)*\cir(#1,0){}
,(5,-25)*\cir(#1,0){}
,(10,-25)*\cir(#1,0){}
,(15,-25)*\cir(#1,0){}
,(20,-25)*\cir(#1,0){}
,(25,-25)*\cir(#1,0){}
,(30,-25)*\cir(#1,0){}
,(35,-25)*\cir(#1,0){}
,(40,-25)*\cir(#1,0){}
,(5,-30)*\cir(#1,0){}
,(10,-30)*\cir(#1,0){}
,(15,-30)*\cir(#1,0){}
,(20,-30)*\cir(#1,0){}
,(25,-30)*\cir(#1,0){}
,(30,-30)*\cir(#1,0){}
,(35,-30)*\cir(#1,0){}
,(40,-30)*\cir(#1,0){}
,(5,-35)*\cir(#1,0){}
,(10,-35)*\cir(#1,0){}
,(15,-35)*\cir(#1,0){}
,(20,-35)*\cir(#1,0){}
,(25,-35)*\cir(#1,0){}
,(30,-35)*\cir(#1,0){}
,(35,-35)*\cir(#1,0){}
,(40,-35)*\cir(#1,0){}
,(5,-40)*\cir(#1,0){}
,(10,-40)*\cir(#1,0){}
,(15,-40)*\cir(#1,0){}
,(20,-40)*\cir(#1,0){}
,(25,-40)*\cir(#1,0){}
,(30,-40)*\cir(#1,0){}
,(35,-40)*\cir(#1,0){}
,(40,-40)*\cir(#1,0){}
#2\end{xy}}}

\newcommand{\rxyzehn}[2]{{\begin{xy} 0;<2mm,0mm>:<0mm,2mm>::0;0,
,(5,-2)*{a}
,(10,-1.8)*{b}
,(15,-2)*{c}
,(20,-2)*{d}
,(25,-2)*{e}
,(30,-2)*{f}
,(35,-2)*{g}
,(40,-2)*{h}
,(45,-2)*{i}
,(50,-2)*{j}
,(2,-5)*{a}
,(2,-10)*{b}
,(2,-15)*{c}
,(2,-20)*{d}
,(2,-25)*{e}
,(2,-30)*{f}
,(2,-35)*{g}
,(2,-40)*{h}
,(2,-45)*{i}
,(2,-50)*{j}
,(5,-5)*\cir(#1,0){}
,(10,-5)*\cir(#1,0){}
,(15,-5)*\cir(#1,0){}
,(20,-5)*\cir(#1,0){}
,(25,-5)*\cir(#1,0){}
,(30,-5)*\cir(#1,0){}
,(35,-5)*\cir(#1,0){}
,(40,-5)*\cir(#1,0){}
,(45,-5)*\cir(#1,0){}
,(50,-5)*\cir(#1,0){}
,(5,-10)*\cir(#1,0){}
,(10,-10)*\cir(#1,0){}
,(15,-10)*\cir(#1,0){}
,(20,-10)*\cir(#1,0){}
,(25,-10)*\cir(#1,0){}
,(30,-10)*\cir(#1,0){}
,(35,-10)*\cir(#1,0){}
,(40,-10)*\cir(#1,0){}
,(45,-10)*\cir(#1,0){}
,(50,-10)*\cir(#1,0){}
,(5,-15)*\cir(#1,0){}
,(10,-15)*\cir(#1,0){}
,(15,-15)*\cir(#1,0){}
,(20,-15)*\cir(#1,0){}
,(25,-15)*\cir(#1,0){}
,(30,-15)*\cir(#1,0){}
,(35,-15)*\cir(#1,0){}
,(40,-15)*\cir(#1,0){}
,(45,-15)*\cir(#1,0){}
,(50,-15)*\cir(#1,0){}
,(5,-20)*\cir(#1,0){}
,(10,-20)*\cir(#1,0){}
,(15,-20)*\cir(#1,0){}
,(20,-20)*\cir(#1,0){}
,(25,-20)*\cir(#1,0){}
,(30,-20)*\cir(#1,0){}
,(35,-20)*\cir(#1,0){}
,(40,-20)*\cir(#1,0){}
,(45,-20)*\cir(#1,0){}
,(50,-20)*\cir(#1,0){}
,(5,-25)*\cir(#1,0){}
,(10,-25)*\cir(#1,0){}
,(15,-25)*\cir(#1,0){}
,(20,-25)*\cir(#1,0){}
,(25,-25)*\cir(#1,0){}
,(30,-25)*\cir(#1,0){}
,(35,-25)*\cir(#1,0){}
,(40,-25)*\cir(#1,0){}
,(45,-25)*\cir(#1,0){}
,(50,-25)*\cir(#1,0){}
,(5,-30)*\cir(#1,0){}
,(10,-30)*\cir(#1,0){}
,(15,-30)*\cir(#1,0){}
,(20,-30)*\cir(#1,0){}
,(25,-30)*\cir(#1,0){}
,(30,-30)*\cir(#1,0){}
,(35,-30)*\cir(#1,0){}
,(40,-30)*\cir(#1,0){}
,(45,-30)*\cir(#1,0){}
,(50,-30)*\cir(#1,0){}
,(5,-35)*\cir(#1,0){}
,(10,-35)*\cir(#1,0){}
,(15,-35)*\cir(#1,0){}
,(20,-35)*\cir(#1,0){}
,(25,-35)*\cir(#1,0){}
,(30,-35)*\cir(#1,0){}
,(35,-35)*\cir(#1,0){}
,(40,-35)*\cir(#1,0){}
,(45,-35)*\cir(#1,0){}
,(50,-35)*\cir(#1,0){}
,(5,-40)*\cir(#1,0){}
,(10,-40)*\cir(#1,0){}
,(15,-40)*\cir(#1,0){}
,(20,-40)*\cir(#1,0){}
,(25,-40)*\cir(#1,0){}
,(30,-40)*\cir(#1,0){}
,(35,-40)*\cir(#1,0){}
,(40,-40)*\cir(#1,0){}
,(45,-40)*\cir(#1,0){}
,(50,-40)*\cir(#1,0){}
,(5,-45)*\cir(#1,0){}
,(10,-45)*\cir(#1,0){}
,(15,-45)*\cir(#1,0){}
,(20,-45)*\cir(#1,0){}
,(25,-45)*\cir(#1,0){}
,(30,-45)*\cir(#1,0){}
,(35,-45)*\cir(#1,0){}
,(40,-45)*\cir(#1,0){}
,(45,-45)*\cir(#1,0){}
,(50,-45)*\cir(#1,0){}
,(5,-50)*\cir(#1,0){}
,(10,-50)*\cir(#1,0){}
,(15,-50)*\cir(#1,0){}
,(20,-50)*\cir(#1,0){}
,(25,-50)*\cir(#1,0){}
,(30,-50)*\cir(#1,0){}
,(35,-50)*\cir(#1,0){}
,(40,-50)*\cir(#1,0){}
,(45,-50)*\cir(#1,0){}
,(50,-50)*\cir(#1,0){}
#2\end{xy}}}

\begin{document}
\title{A Dark Sector Extension of the Almost-Commutative Standard Model}
\author{Christoph A.~Stephan \footnote{Institut f\"ur Mathematik, Universit\"at Potsdam, Am Neuen Palais 10, Potsdam, Germany}   }

\maketitle

\begin{abstract}
\noindent
We consider an extension of the Standard Model within the frame work of Noncommutative Geometry. The 
model is based on an older model [St09] which extends the Standard Model by new fermions, a new $U(1)$-gauge
group and, crucially, a new scalar field which couples to the Higgs field. This new scalar field allows to lower the
mass of the Higgs mass from $\sim 170$ GeV, as predicted by the Spectral Action for the Standard Model,
to a value of $120-130$ GeV. The short-coming of the previous model lay in its inability to meet all the constraints 
on the gauge couplings implied by the Spectral Action. These shortcomings are cured in the present model which 
also features a ``dark sector'' containing fermions and scalar particles.
\end{abstract}


\section{Introduction}

Noncommutative Geometry has  in the last two decades   proved to be of considerable 
interest for particle physics. The construction of the Standard Model of particle physics in terms of spectral triples  
\cite{C94,CM08} provides deep insights into the geometric nature of high energy physics. In conjunction with the 
Spectral Action \cite{CC97} on obtains a highly predictive, mathematically sound foundation of particle physics.
For readers interested in an introduction to the field we recommend \cite{Sc05a} and \cite{DS12}.
\medskip

\noindent
The geometrical basis for the construction of particle models is provided by almost-commutative geometries
consisting of a spectral triple on a compact Riemannian manifold to model space(time) and an internal space constructed
from a matrix algebra. The geometry and the dynamics of the models are encoded in a generalised Dirac operator which
comprises the Dirac operator on the manifold, covariant derivatives with respect to the gauge group of the model
and scalar fields (and Dirac or Majorana masses) with their Yukawa matrices.  The particle content of the model and the 
interactions are fixed by the matrix algebra and its representation on the spinor space. The dynamical part of the theory is
given by the Dirac action and the Spectral Action. Interpreting the Dirac Action and the Spectral Action as an effective action,
valid at some cut-off energy,  imposes further constraints on the  model's coupling constants. These constraints make the Spectral
Action highly predictive, namely for the pure Standard Model one predicts the mass of the Higgs boson to be $\sim 170$ GeV, 
\cite{CC97}. This value has recently been shown to be too high \cite{ATLAS12,CMS12}, the Higgs has a mass of $\sim 125$ GeV.
\medskip

\noindent
Although the Standard Model takes a prominent place \cite{ISS04,JS08,CC08} 
within the possible almost-commutative geometries one can go further and construct models beyond 
the Standard Model. Finding extension within the almost-commutative framework of the classical algebra and with respect
to the classical axioms \cite{C96} is dating back to the beginnings of the field, \cite{PS97,PSS99,SZ01}. In these extensions
the Standard Model algebra $\mathcal{A}_{SM} = \mathbb{C} \oplus \mathbb{H} \oplus M_3(\mathbb{C})$ or the
Dirac operator were only mildly enlarged to include for example lepto-quarks.  More recently the techniques from the
classification scheme developed in \cite{ISS04} were used to enlarge the models further, \cite{St06a,St07,SS07,St09}.
Here the model in \cite{St09} will be of central interest, since it predicted approximately the correct Higgs mass.
\medskip

\noindent
In the case of finite spectral triples of KO-dimension six  \cite{B07,C06,CCM07}
another classification  leads to more general versions of the Standard Model algebra \cite{CC08,CC10} under some extra assumptions. 
Considering the first order axiom, see Appendix B, as being dynamically imposed on the spectral triple one finds a Pati-Salam
type model  \cite{CCS13}. From the same geometrical basis one can promote the Majorana mass of the neutrinos to a scalar field
\cite{CC12,DLM13} which allows to lower the Higgs mass to its experimental value.
Another recent line of research which is aiming at supersymmetric extensions of the Standard Model has been started in \cite{BS11,BS12}.
The inclusion of grand unified theories seems to necessitate non-associative spectral triples \cite{W97,FB13} 
and has recently gained interest again. One can furthermore include torsion  \cite{HPS10,PS12} and also impose scale invariance 
introducing dilaton fields \cite{CC06,AKL11}.
\medskip

\noindent
For the present paper we concentrate on a variation of the model in  \cite{St09}. The original model contains a new scalar field mixing with the 
Standard Model Higgs.  Unfortunately it does not allow to meet all the constraints
imposed by the Spectral Action. It was a long held believe \cite{LS01} that fluctuations of the Dirac operator by 
centrally extended lifts allow to eliminate such constraints for the gauge couplings of abelian gauge groups.  This claim is not
true and so the problem of the non-matching triangle for the high energy values of the gauge couplings of the Standard Models
prevails. We will show in this paper how to construct a model that overcomes these short comings by enlarging the particle content
while respecting the axioms of noncommutative geometry. The model is compatible with the measured value of the Higgs mass, 
has a matching of the running gauge couplings consistent with the constraints from the Spectral Action and ensures that 
the potential of the scalar fields remains stable up to the Planck mass.   Furthermore the new particles may have interesting 
phenomenological implications, perhaps as candidates for dark matter.
\medskip

\noindent
In section \ref{SA} we will give a short introduction into the differential geometric setting of generalised Dirac operators, in particular 
those of Chamseddine-Connes type, and the Spectral Action. Section \ref{M} gives  details of the model covering
the gauge group, its representation on the fermionic Hilbert space and the Lagrangian of the model. Furthermore the high energy
boundary conditions of the Spectral action are computed. The numerical analysis
for a point in the parameter space of the model is carried out in section \ref{NE}. Appendix A provides details on the normalisation
conventions of the Yang-Mills sector and in Appendix B the details of the underlying spectral triple are discussed. The reader mainly
interested in the physical model can concentrate on the sections \ref{M} and \ref{NE}.

\section{The Spectral Action \label{SA}}
From the differential geometric point of view one can consider the Dirac operator of an almost-commutative spectral triple as 
a generalised twisted Dirac operator on a Riemannian manifold $M$. So any particle model for which the fermionic action is expressible in 
terms of such a generalised Dirac operator qualifies to be investigated from the spectral point of view, i.e. it may be worthwhile 
to extend the Spectral Action principle to particle models that are not necessarily based in noncommutative geometry.
In these models of particle physics the matter content is encoded in a Hermitian vector bundle $\HH\to M$ equipped with a connection $\nabla^\HH$.
The Levi-Civita connection $\nabla^g$  induces a connection on the spinor bundle $\Sigma M$, which we denote again by $\nabla^g$.
A symmetric Dirac operator  on the twisted bundle $\mathcal{E}=\Sigma M\otimes \HH$ is defined
in terms of the Levi-Civita connection $\nabla^g$ and $\nabla^\HH$:
\begin{equation}
D_\HH(\psi\otimes\chi)=\sum_{i=1}^4\left((e_i\cdot \nabla^g_{e_i}\psi)\otimes \chi+
(e_i\cdot \psi)\otimes (\nabla^\HH_{e_i}\chi) \right)
\label{def_twist_Dirac}
\end{equation}
for any positively oriented orthonormal frame $e_1,\ldots, e_4$, any section $\psi$ of $\Sigma M$ and any section $\chi$ of $\HH$.
For the twisted Dirac operator $D_\HH$ we get
\begin{equation}
D_\HH^2 (\psi\otimes\chi)= \Delta^\nabla (\psi\otimes\chi) + \tfrac14 R^g \psi \otimes \chi-
\tfrac12\,\sum_{i\ne j}(e_i\cdot e_j\cdot\psi)\otimes\Omega_{ij}^\HH\chi
\label{eq_twist_Bochner}
\end{equation}
where $\Delta^\nabla$ is the Laplacian associated to the  connection
\[
\nabla=\nabla^g \otimes \id_\HH  + \id_{\Sigma M}\otimes \nabla^\HH,
\]
$R^g$ is the scalar curvature of the manifold and 
$\Omega_{ij}^\HH=\nabla^\HH_{e_i}\nabla^\HH_{e_j}-\nabla^\HH_{e_j}\nabla^\HH_{e_i}-\nabla^\HH_{[e_i,e_j]}$ is the curvature of $\nabla^\HH$.
The curvatures of $\nabla$, $\nabla^g$ and $\nabla^\HH$ are related as:
\begin{equation}
\Omega_{ij}^\nabla(\psi\otimes\chi)=
(\Omega_{ij}^g\psi)\otimes\chi +\psi\otimes(\Omega_{ij}^\HH\chi).
\label{eq_curvature-addition}
\end{equation}
\medskip

\noindent
The fluctuated Dirac operator can then be expressed in terms of a selfadjoint endomorphism $\Phi$ on $\HH$ 
added to the twisted Dirac operator,
\bb
D_\Phi(\psi\otimes\chi)=D_\HH(\psi\otimes\chi) +(\omega^g\cdot\psi)\otimes (\Phi\,\chi)
\label{CCDiracOp}
\ee
where $\omega^g$ denotes the volume element. The {\it Higgs endomorphism} $\Phi$ encodes the
scalar fields, the Yukawa masses or couplings and Dirac mass terms. 
For $D_\Phi$ one has the Lichnerowicz formula
\begin{equation}
D_\Phi^2 (\psi\otimes\chi)= \Delta^\nabla(\psi\otimes\chi)-E_\Phi(\psi\otimes\chi)
\label{CCBochner}
\end{equation}
where the potential $E_\Phi$ is given by
\bb
E_\Phi(\psi\otimes\chi) =
- \tfrac14 R^g \psi\otimes \chi+
\tfrac12\,\sum_{i\ne j}(e_i\cdot e_j\cdot\psi)\otimes\Omega_{ij}^\HH\chi
+\sum_i (\omega^g\cdot e_i\cdot\psi)\otimes \big[\nabla^\HH_{e_i},\Phi \big]\chi
-\psi\otimes(\Phi^2\,\chi) .
\label{CCBochner_potential}
\ee
The {\it Spectral Action} $\mathcal{S}_{CC}$ for $D_\Phi$ is defined  as the number of eigenvalues in an interval $[-\Lambda,+\Lambda]$ for some cut-off
energy $\Lambda$. The Spectral Action is a counting function and can be expressed as $\mathcal{S}_{CC}(D_\Phi) = \tr_\HH [f(D_\Phi^2 /\Lambda^2)]$ with
a suitable cut-off function $f : \R^+ \to \R^+$ which has support in the interval $[0,+1]$ and is constant near the origin.
 Performing a Laplace transformation and a heat kernel expansion \cite{CC97,NVW02} one finds the asymptotic expression
\bb
\mathcal{S}_{CC}(D_\Phi) =  f_0 \Lambda^4 a_0(D_\Phi^2) + f_2 \Lambda^2 a_2(D_\Phi^2)  + f_4 a_4(D_\Phi^2) + \mathcal{O}(\Lambda^{-\infty})
\label{SpectralAction}
\ee
as $\Lambda \to \infty$. Here $f_0$, $f_2$ and $f_4$ are the first moments of the cut-off function $f$. We will consider them as free parameters to be determined
by experiment. 
Using Gilkey's general formulas \cite{G95}  one obtains for $D_\Phi^2$ the following Seeley-DeWitt coefficients \cite{CC97,IKS97,CCM07,CC10}

\bb
a_0(D_\Phi^2) &=& \tfrac{1}{4\,\pi^2} \,\rk (\HH)\,\vol(M) ,
\label{a0} \\
a_2(D_\Phi^2) &=& -\tfrac{\rk (\HH)}{48\,\pi^2}\,\int_M R^g \, dx
-\tfrac{1}{4\,\pi^2}\int_M\tr_\HH(\Phi^2)\,dx, 
\label{a2} \\
a_4(D_\Phi^2) &=& \tfrac{11\,\rk(\HH)}{720}\,\chi(M)
-\tfrac{\rk(\HH)}{320\,\pi^2}\int_M\|C^g\|^2\,dx -\tfrac{1}{24\,\pi^2}\int_M\tr_\HH(\Omega^\HH\Omega^\HH)\,dx 
\nonumber \\
&&\quad +\tfrac{1}{8\,\pi^2}\int_M\left( \tr_\HH([\nabla^\HH,\Phi]^2)+\tr_\HH(\Phi^4)
+\tfrac{1}{6}\, R^g \tr_\HH(\Phi^2)\right)\,dx
\label{a4}
\ee
with the abbreviations $[\nabla^\HH,\Phi]^2=\sum_i [\nabla^\HH_{e_i},\Phi][\nabla^\HH_{e_i},\Phi] $ and $\Omega^\HH\Omega^\HH=\sum_{i,j}\Omega^\HH_{ij}\Omega^\HH_{ij}$.
\medskip

\noindent
Assume now that the  principle fibre bundle of the particle model  has the structure group 
$G := G_1\times \cdots \times G_m$ with subgroups $G_s$ either equal to $U(1)$ or $SU(n_s)$ with connection one-forms  $\omega^s$ and curvature
two-forms $\Omega^s$. Then the Yang-Mills action is given by the normalisation
\bb
-\frac{f_4}{24 \pi^2} \int_M\tr_\HH(\Omega^\HH\Omega^\HH)\,dx   \stackrel{!}{=} 
\frac{1}{4 g_1^2} \int_M \sum_{a,i,j} (\Omega^1)^a_{ij} (\Omega^1)^a_{ij})\,dx  \; + \cdots + \;  
\frac{1}{4 g_s^2} \int_M \sum_{a',i,j} (\Omega^s)^{a'}_{ij} (\Omega^s)^{a'}_{ij})\,dx, 
\label{YangMillsnorm}
\ee
where $g_s$ denotes the respective gauge coupling. The details for the choice of the Lie algebra basis are given in Appendix A. 
Furthermore  assume that $\Phi$ encodes $r$ complex scalar (multiplet) fields $\varphi^1,\dots \varphi^r$. Then the scalar part of the action
is normalised to
\bb
&&\int_M \left( - \frac{f_2 \, \Lambda^2}{4 \pi^2}  \tr_\HH(\Phi^2)  + \frac{f_4}{8 \pi^2} \tr_\HH([\nabla^\HH,\Phi]^2)
+  \frac{f_4}{8 \pi^2} \tr_\HH(\Phi^4) \right) \; dx
\nonumber \\
&\stackrel{!}{=}&  \int_M \sum_j \left( |\nabla^j \varphi^j|^2  -  \mu_j^2 |\varphi^j|^2  +  \frac{\lambda_j}{6} |\varphi^j|^4 +
 \sum_{i < j} \frac{\lambda_{r+i}}{3} |\varphi^i|^2 |\varphi^j|^2\right) \; dx + ({\rm mass \; terms})^4 \vol(M)
\label{Scalarnormalisation} 
\ee
where the mass terms can be absorbed into the cosmological term which is identified with $f_0 \Lambda^4 a_0(D_\Phi^2)$. 
The scalar fields $\tilde \varphi^i$ obtained from the fluctuations of the Dirac operator have mass dimension zero
while the Yukawa mass matrices $M_{\chi}$ have mass dimension one. The dynamical
term in \eqref{Scalarnormalisation} allows to determine the normalisation for the physical scalar fields $\varphi^i$ with
mass dimension one in  terms of the Yukawa mass matrices $\M_{\chi_i}$ and scalar fields $\tilde \varphi^i$.
Imposing the standard normalisation of scalar-spinor interaction terms in the Dirac action at $\Lambda$
eliminates the Yukawa mass matrices $\M_{\chi_i}$  in favour of the Yukawa coupling matrices $g_{\chi_i}$ with mass dimension
zero:
\bb
\langle \chi, \Phi(\M_{\chi_i},\tilde \varphi^i) \, \chi \rangle \stackrel{!}{=}   \langle \chi, \Phi(g_{\chi_i}, \varphi^i) \, \chi \rangle.
\label{Diracnormalisation}
\ee
If the Higgs endomorphism of the Dirac operator is constructed directly in terms of Yukawa couplings and scalar fields with mass dimension one it is not necessary to
invoke the normalisation relation  {\eqref{Diracnormalisation}.  
\medskip

\noindent 
The  standard normalisation of the Einstein-Hilbert action and for the coupling of the scalar curvature to the
scalar fields implies 
\bb
-\frac{f_2 \, \Lambda^2 \, \rk(\HH)}{48 \pi^2} \int_M R^g \; dx \; +  \; \frac{f_0}{8 \pi^2}  \int_M R^g \, \tr_\HH(\Phi^2) \; dx
\stackrel{!}{=} -\frac{m_p^2}{16 \pi} \int_M R^g \; dx \; + \; \int_M R^g \, \sum_i \beta_i \, |\varphi^i|^2 \; dx
\label{Einsteinnormalisation}
\ee
where the $\beta_i$ are real parameters.

\noindent
The physical normalisation relations \eqref{YangMillsnorm}, \eqref{Scalarnormalisation}, \eqref{Diracnormalisation} and \eqref{Einsteinnormalisation} 
imply relations among the gauge couplings, the quartic scalar couplings and the
Yukawa couplings \cite{CC97,T03}. Since these relations are not stable under the renormalisation group flow one interprets the Spectral Action as
an effective action valid at a cut-off scale $\Lambda$. Ideally this cut-off scale can be chosen such that all relations among the couplings
are fulfilled at $\Lambda$ while the low energy values, say at the Z-boson mass, agree with experimental data.

\section{The Model \label{M}}

The model we are investigating here can be either formulated directly in terms of the generalised twisted Dirac operator \eqref{CCDiracOp} 
or it can be constructed from a spectral triple, see Appendix B. The internal Hilbert space $\HH$ of the model 
extends the Standard Model Hilbert space \cite{CC97} by $N$ generations of vector-like $X_1$-particles,
chiral $X_2$- and $X_3$-particles and vectorlike $V_{c/w}$-particles.  The structure group of the Standard Model is enlarged by
an extra $U(1)_X$ subgroup, so the total group is $G=U(1)_Y \times SU(2)_w \times SU(3)_c \times U(1)_X$. 
The Standard Model particles and the $V_{c/w}$-particles are neutral with respect to the $U(1)_X$ subgroup while the $X$-particles 
are neutral with respect to the  Standard Model subgroup $G_{SM}=U(1)_Y \times SU(2)_w \times SU(3)_c $. 
Furthermore the model contains two scalar fields: a scalar field in the Standard Model Higgs representation and a new scalar field 
carrying only a $U(1)_X$ charge.
\medskip

\noindent
The Hilbert spaces of the fermions and the scalar fields expressed in terms of their detailed representations of $G$ are written out in detail in 
the following list. The subscripts indicate the particle species in question, e.g. $q$, $\ell$ for quarks and leptons, $u$, $d$ for 
up- or down-type quarks and  $e$, $\nu$ for electron- or neutrino-type leptons. The superscript $p$ indicates that we restrict to the particle 
sector for brevity. 
\bb
&&\HH_{q,l}^p\oplus \HH_{\ell, l}^p  = \bigoplus_1^3 [(+\tfrac16,2,3,0) \oplus (-\tfrac12,2,1,0)] 
\nonumber \\
&&\HH_{u,r}^p\oplus \HH_{d,r}^p\oplus  \HH_{e, r}^p \oplus \HH_{\nu ,r}^p  = \bigoplus_1^3 [(+\tfrac23,1,3,0) \oplus (-\tfrac13,1,3,0) \oplus (-1,1,1,0)\oplus (0,1,1,0)]
\nonumber \\
&& \HH_{X_1,l}^p \oplus \HH_{X_2,l}^p \oplus \HH_{X_3,l}^p = \bigoplus_1^N [(0,1,1,+1) \oplus (0,1,1,+1) \oplus (0,1,1,0)]
\nonumber \\
&& \HH_{X_1,r}^p \oplus \HH_{X_2,r}^p \oplus \HH_{X_3,r}^p = \bigoplus_1^N [(0,1,1,+1) \oplus (0,1,1,0) \oplus (0,1,1,+1)]
\nonumber \\
&& \HH_{V_c,l}^p \oplus \HH_{V_w,l}^p = \HH_{V_c,r}^p \oplus \HH_{V_w,r}^p  = \bigoplus_1^N [(-\tfrac13,1,\bar 3,0) \oplus (0,\bar 2,1,0)]
\label{repsfermions} \\
&& \HH_{H} = ( -\tfrac12,2,1,0), \quad \HH_{\varphi} = (0,1,1,-1)
\ee
Here we chose the standard normalisations of \cite{MV83,MV84,MV85} for the representation, i.e. the charges, of the abelian subgroups.
The details for the Higgs endomorphism of the Dirac operator are given in Appendix B. 
\medskip

\noindent
Let us now summarise the terms that will comprise the Lagrangian \eqref{totalLagrangian} of the model. Readers not interested in the 
computational details of the Spectral Action or the construction of the spectral triple can take the Lagrangian in this
subsection as a definition and as a starting point for phenomenological investigations. 
Together with the boundary conditions at the cut-off scale, \eqref{AllUatLambda}, \eqref{YukawasatLambda}, \eqref{lambda1atLambda},
\eqref{lambda2atLambda} and \eqref{lambda3atLambda}, the Lagrangian \eqref{totalLagrangian} constitutes 
the main physical content of the model as an effective field theory.
\medskip

\noindent
Taking the Dirac inner product 
$\langle \psi, D_\Phi \, \psi \rangle$ as defined in  \cite{CCM07} we can decompose it into the fermionic subspaces in \eqref{repsfermions}.
In this way we get the fermionic Lagrangian of the Standard Model in its euclidean formulation, the fermionic Lagrangian for the $X$-particles, 
\bb
\mathcal{L}_{\nu X_1} &=&  \sum_{i=1}^3 \sum_{j=1}^N \; \left( (g_{\nu X_1})_{ij} \, \psi_{\nu^i,r}^* \, \gamma_5 \, \varphi \, \psi_{X_1^j,l} \, +  \, {\rm h.c.} \right)
\nonumber \\
\mathcal{L}_{X_1} &=& \sum_{i=1}^N \; ( \psi_{X_1^i,l}^* , \psi_{X_1^i,r}^*) D_{X_1^i}  \pp{\psi_{X_1^i,l} \\ \psi_{X_1^i,r} }
\; + \; \sum_{i,j=1}^N \; \left( (\M_{X_1})_{ij} \, \psi_{X_1^i,l}^* \, \gamma_5 \, \psi_{X_1^j,r} \, +  \, {\rm h.c.} \right)
\nonumber \\
\mathcal{L}_{X_2} &=& \sum_{i=1}^N \; ( \psi_{X_2^i,l}^* , \psi_{X_2^i,r}^*) D_{X_2^i}   \pp{\psi_{X_2^i,l} \\ \psi_{X_2^i,r} }
\; + \; \sum_{i,j=1}^N \; \left( (g_{X_2})_{ij} \, \psi_{X_2^i,l}^* \, \gamma_5 \, \varphi \, \psi_{X_2^j,r} \, +  \, {\rm h.c.} \right)
\nonumber \\
\mathcal{L}_{X_3} &=& \sum_{i=1}^N \; ( \psi_{X_3^i,l}^* , \psi_{X_3^i,r}^*) D_{X_3^i}  \pp{\psi_{X_3^i,l} \\ \psi_{X_3^i,r} }
\; + \; \sum_{i,j=1}^N \; \left( (g_{X_3})_{ij} \, \psi_{X_3^i,l}^* \, \gamma_5 \, \bar \varphi \, \psi_{X_3^j,r} \, +  \, {\rm h.c.} \right),
\label{XLagrangian}
\ee
and the fermionic Lagrangian of the $V_c$- and the $V_w$- particles
\bb
\mathcal{L}_{V_c} &=& \sum_{i=1}^N \; ( \psi_{V_c^i,l}^* , \psi_{V_c^i,r}^*) D_{V_c^i}  \pp{\psi_{V_c^i,l} \\ \psi_{V_c^i,r} }
\; + \; \sum_{i,j=1}^N \; \left( (\M_{V_c})_{ij} \, \psi_{V_c^i,l}^* \, \gamma_5 \, \psi_{V_c^j,r} \, +  \, {\rm h.c.} \right)
\nonumber \\
\mathcal{L}_{V_w} &=& \sum_{i=1}^N \; ( \psi_{V_w^i,l}^* , \psi_{V_w^i,r}^*) D_{V_w^i}  \pp{\psi_{V_w^i,l} \\ \psi_{V_w^i,r} }
\; + \; \sum_{i,j=1}^N \; \left( (\M_{V_w})_{ij} \, \psi_{V_w^i,l}^* \, \gamma_5 \, \psi_{V_w^j,r} \, +  \, {\rm h.c.} \right)
\label{VLagrangian}
\ee
The Dirac operators of the form  $D_X$ and $D_V$ in \eqref{XLagrangian} and \eqref{VLagrangian} are the  twisted Dirac operators
with gauge covariant derivatives according to the representations in \eqref{repsfermions}. Observe that the mass terms of the
$X_1$-particles and the $V_{c/w}$-particles are given by Dirac mass matrices since their charges are left-right symmetric. 
The mass terms of the $X_{2/3}$-particles are induced by the new scalar field $\varphi$, where we have taken the liberty to 
normalise the scalar field to mass dimension one and the Yukawa couplings to mass dimension zero. If the Dirac operator is
obtained from fluctuations, see Appendix B, then the mass dimension of the scalar fields is zero and we have Yukawa mass matrices.
The formal justification for passing from one normalisation to the other will be given in the next section.
\medskip

\noindent
The Yang-Mills Lagrangian for the  Standard Model subgroup $G_{SM}$ takes again its usual form  and for the $U(1)_X$ subgroup we add
\bb
\mathcal{L}_{U(1)_X} = \frac{1}{4 g_4^2}  \sum_{i,j=1}^4 (B^X)_{ij} (B^X)_{ij}, 
\ee
where $B^X$ is the field strength tensor for the $U(1)_X$-covariant derivative. For details on the normalisation see Appendix A.
Finally the Lagrangian of the scalar fields  
\bb
\mathcal{L}_{H,\varphi} =  |\nabla^{H} H |^2  +  |\nabla^{\varphi} \varphi|^2 - \mu_H^2 |H|^2 - \mu_\varphi^2 |\varphi|^2 + 
\frac{\lambda_1}{6} |H|^4 + \frac{\lambda_2}{6} |\varphi|^4 + \frac{\lambda_3}{3} |H|^2 \, |\varphi|^2 
\label{ScalarLagrangian}
\ee
contains the dynamical terms as well as the the symmetry breaking potential with interaction term of the $H$-field and the $\varphi$-field.
The normalisation is again chosen according to \cite{MV83,MV84,MV85} in order to obtain Lagrangians of the form
$(1/2) (\partial \varphi)^2 -(\mu^2/2) \varphi^2 + (\lambda/24) \varphi^4+\dots$ for the real valued fields.
Putting everything together the Dirac inner product  $\langle \psi, D_\Phi \, \psi \rangle$  and the Spectral Action \eqref{SpectralAction} 
provide the Lagrangian
\bb
\mathcal{L}_{full} = \mathcal{L}_{SM,f} + \mathcal{L}_{SM,YM} + \mathcal{L}_{U(1)_X} + \mathcal{L}_{H,\varphi} + \mathcal{L}_{\nu X_1} +\mathcal{L}_{X_1}
+\mathcal{L}_{X_2} + \mathcal{L}_{X_3} + \mathcal{L}_{V_c} + \mathcal{L}_{V_w}.
\label{totalLagrangian}
\ee

\subsection{Dimensionless parameters}

The Spectral Action \eqref{SpectralAction} does not only supply the  bosonic part of the Lagrangian \eqref{totalLagrangian} but also
relations among the free parameters of the model. These relations serve as boundary conditions at the cut-off scale $\Lambda$
for the renormalisation group flow. Let us start with the computation of the Yang-Mills terms in \eqref{SpectralAction} which 
are encoded in the $\tr_\HH (\Omega^\HH \Omega^\HH)$-term in the fourth Seeley-DeWitt coefficient $a_4(D^2_\Phi)$ in \eqref{a4}.
Here the trace $\tr_\HH$ is taken over the whole inner Hilbert space of the fermions.

\paragraph{$U(1)_Y$-coupling relation}

We first calculate the total sum of the squares of the $U(1)_Y$ hypercharges $Q_Y^2$, taking into account
their multiplicities according to the dimension of the respective fermionic subspaces consisting of the Standard Model particles and
the $V_{c}$-particles.
\bb
Q_Y^2 := \overbrace{2\cdot 3 \cdot 3\cdot [2\cdot (\tfrac16)^2 + (\tfrac23)^2 + (-\tfrac13)^2] }^{\rm quarks} + \overbrace{2\cdot  3 \cdot [2\cdot (-\tfrac12)^2 + (-1)^2]}^{\rm leptons} 
+ \overbrace{2\cdot  N \cdot 3 \cdot 2 \cdot(-\tfrac16)^2}^{V_c-\rm particles}
\eee
The trace of the squared Yang-Mills curvature terms can then be computed for each subgroup of $G$ separately, see Appendix A.
Inserting the relevant term $a_4(D^2_\Phi)$ into the Spectral Action \eqref{SpectralAction} yields after normalising 
\bb
-\frac{f_4}{24 \pi^2} \; Q_Y^2 \, \int_M \sum_{i,j}(\Omega^Y)_{ij}(\Omega^Y)_{ij}\,dx 
&=& \frac{f_4}{24 \pi^2} \frac14 (80 + \tfrac43 N) \int_M \sum_{i,j} B_{ij} B_{ij} \,dx
\nonumber \\
&\stackrel{!}{=}& \frac{1}{4 g_1^2} \int_M \sum_{i,j} B_{ij} B_{ij} \,dx
\eee 
the relation 
\bb
g_1^2 = \frac{24 \pi^2}{f_4} (80+ \tfrac43 N)^{-1}
\label{U1atLambda}
\ee
between the gauge coupling $g_1$, the fourth moment of the cut-off function $f_4$ and $Q_Y^2$.

\paragraph{$SU(2)_w$-coupling relation}

Proceeding as in the previous paragraph we calculate the trace of the $SU(2)_w$ curvature terms in $a_4(D^2_\Phi)$:
\bb
-\frac{f_4}{24 \pi^2} \left( \overbrace{2\cdot 3 \cdot 3}^{\rm quarks} + \overbrace{2\cdot 3 }^{\rm leptons} + \overbrace{2 \cdot N \cdot 2}^{V_w-\rm part.}  \right) \int_M \tr_\HH(\Omega^w\Omega^w)\,dx &=& \frac{f_4}{48 \pi^2} (24 + 4 N) \int_M \sum_{a,i,j} W^a_{ij} W^a_{ij} \,dx
\nonumber \\
& \stackrel{!}{=}& \frac{1}{4 g_2^2} \int_M \sum_{a,i,j} W^a_{ij} W^a_{ij} \,dx
\eee
The normalisation 
\bb
g_2^2 = \frac{24 \pi^2}{f_4} (48+ 8 N)^{-1}
\label{SU2atLambda}
\ee
yields a relation between the gauge coupling $g_2$ and the fourth moment of the cut-off function $f_4$.

\paragraph{$SU(3)_c$-coupling relation}
Repeating the steps from the $SU(2)_w$ curvature terms in $a_4(D^2_\Phi)$ for the $SU(3)_c$ curvature terms,
\bb
-\frac{f_4}{24 \pi^2} \left( \overbrace{2\cdot 3 \cdot 4}^{\rm quarks} + \overbrace{2 \cdot N \cdot 2}^{V_c-\rm part.}  \right) \int_M \tr_\HH(\Omega^c\Omega^c)\,dx &=& \frac{f_4}{48 \pi^2} (24 + 4 N) \int_M \sum_{a,i,j} G^a_{ij} G^a_{ij} \,dx
\nonumber \\
& \stackrel{!}{=}& \frac{1}{4 g_3^2} \int_M \sum_{a,i,j} G^a_{ij} G^a_{ij} \,dx
\eee
we find the relation 
\bb
g_3^2 = \frac{24 \pi^2}{f_4} (48+ 8 N)^{-1}.
\label{SU3atLambda}
\ee
Here we would like to note the deviations from the relations that appear in the Standard Model case. We see the additional factors
proportional to $N$ in \eqref{U1atLambda}, \eqref{SU2atLambda} and \eqref{SU3atLambda} which have their origin in the $V_c$-particles
and the $V_w$-particles. Similar variations of such relations have already been considered in \cite{St07,SS07,St09}.

\paragraph{$U(1)_X$-coupling relation}

For the total sum of the squared $U(1)_Y$ charges of the $X$-particles we find 
\bb
Q_X^2 :=  \overbrace{2\cdot N \cdot 2 \cdot 1^2 }^{X_1\rm -particles} 
+ \overbrace{2\cdot N \cdot 1^2}^{X_2\rm-part.} + \overbrace{2\cdot N \cdot 1^2}^{X_3\rm-part.}. 
\eee
This allows us to write the trace of the $U(1)_X$ curvatures squared in $a_4(D_\Phi^2)$ as
\bb
-\frac{f_4}{24 \pi^2}\; Q_X^2 \; \int_M \sum_{i,j}(\Omega^X)_{ij}(\Omega^X)_{ij}\,dx 
&=& \frac{f_4}{24 \pi^2} \frac14 \, 32 N \int_M \sum_{i,j} B^X_{ij} B^X_{ij} \,dx
\nonumber \\
& \stackrel{!}{=}& \frac{1}{4 g_4^2} \int_M \sum_{i,j} B^X_{ij} B^X_{ij} \,dx
\eee
and we obtain the relation
\bb
g_4^2 = \frac{24 \pi^2}{f_4} \frac{1}{32 N}.
\label{UXatLambda}
\ee

\paragraph{Gauge coupling relations}

All four relations \eqref{U1atLambda}, \eqref{SU2atLambda}, \eqref{SU3atLambda} and \eqref{UXatLambda} depend linearly on 
the fourth moment of the cut-off function $f_4$. Therefore we can eliminate $f_4$ to obtain the final relations among $g_1$, $g_2$,
$g_3$ and $g_4$ at the cut-off scale $\Lambda$:
\bb
\boxed{
\sqrt{\frac{80 + \tfrac43 N}{48 + 8 N}} \; g_1(\Lambda) = g_2(\Lambda)=g_3(\Lambda) = \sqrt{\frac{32 N}{48 + 8 N}} \; g_4(\Lambda) 
}
\label{AllUatLambda}
\ee
We notice again the deviation of these relations compared to the case of the Standard Model, see \cite{CC97}. Yet the 
relation $ g_2(\Lambda)=g_3(\Lambda) $ remains unchanged, although the actual value of $\Lambda$ may still deviate 
from the Standard Model result, since the $V_{c/w}$ are charged under the Standard Model subgroup $G_{SM}$ and can
therefore change the running of the couplings under renormalisation group flow.

\paragraph{Remark:} We would like to point out that the relations for the abelian subgroups cannot be normalised away by choosing different numerical 
values of the $U(1)$-charges in the central extension, see Appendix B. This possibility of reducing the number of boundary conditions
was implied in \cite{LS01}. It was used in many succeeding publications, including the publications of the author, see \cite{St07,St09}, etc. 
Yet, the conclusion that the free choice of the $U(1)$-charges implies that the boundary condition of the respective gauge coupling is empty does 
not hold, since only the product of the (squared) $U(1)$-charges and the gauge coupling is a measurable physical quantity.  But this 
product can be fixed, for example by a measurement of the numerical value or by boundary conditions such as \eqref{AllUatLambda}. Normalising
the value of the $U(1)$-charges forces automatically a reciprocal normalisation of the gauge coupling resulting in the same value for the product.
Therefore the boundary conditions of {\it all} gauge couplings need to be taken into account.

\paragraph{Yukawa coupling relations}

Next we will find  relations similar to \eqref{AllUatLambda} for the traces of the squared Yukawa coupling matrices $Y_2$ and $Y_X$, for the
definitions see \eqref{squaredsums}. We compute the normalisation of the kinetic part of  $a_4(D_\Phi^2)$ for the Higgs endomorphism in
 in terms of 
the scalar fields $\tilde H$ and $\tilde \varphi$ with mass dimension zero. Inserting this into the Spectral Action \eqref{SpectralAction} and equating this part of the 
Lagrangian with the standard kinetic Lagrangian for the mass dimension one fields $H$ and $\varphi$,
\bb
\tfrac{1}{8\,\pi^2}\int_M  \tr_\HH([\nabla^\HH,\Phi]^2 ) \,dx
&=& \tfrac{1}{8\,\pi^2}\int_M\left( 4 \Y_2 \,\tr ([\nabla^{\tilde H}, \tilde{H}]^2) + 4 \Y_X \, |\nabla^{\tilde \varphi}\tilde \varphi|^2  \right)  \,dx
\nonumber \\
&\stackrel{!}{=}& \int_M\left(   |\nabla^{H} H |^2  +  |\nabla^{\varphi} \varphi|^2  \right)  \,dx
\ee
yields the normalisation in terms of the traces of the squared Yukawa mass matrices \eqref{squaredmasssums}:
\bb
\tr(\tilde H^*\tilde H) = \frac{2 \pi^2}{f_4} \frac{1}{\Y_2} \, |H|^2, \qquad |\tilde \varphi |^2 = \frac{2 \pi^2}{f_4} \frac{1}{\Y_X} |\varphi|^2.
\label{Yukawamassrelations}
\ee
If the Higgs endomorphism had already been given in terms of scalar fields of mass dimension one and Yukawa coupling matrices, then the
relations \eqref{Yukawamassrelations} would immediately imply the desired boundary conditions \eqref{YukawasatLambda} as was first noted in \cite{T03}. In the present case we need a 
second relation to eliminate the spurious mass  scale. This relation is provided by the Dirac action. We write the Higgs endomorphism as a function of the 
Yukawa mass matrices and the scalar fields of mass dimension one, $\Phi = \Phi(\M_{SM},\M_X, \tilde H, \tilde \varphi)$, and impose equality of the Dirac action with the
equivalent normalisation with respect to the Yukawa couplings and mass dimension zero scalar fields. Assuming the endomorphism $\Phi$ is linear in
the Yukawa matrices and the scalar fields we find that 
\bb
\langle \chi, \Phi (\M_{SM},\M_X, \tilde H, \tilde \varphi)  \chi \rangle 
&=&  \sqrt{\frac{2 \pi^2}{f_4}}\; \langle \chi, \Phi (\frac{\M_{SM}}{\sqrt{\Y_2}},\frac{\M_X}{\sqrt{\Y_X}}, H, \varphi)  \chi \rangle 
\nonumber \\
&=& \sqrt{ \frac{2 \pi^2}{f_4}}\; \langle \chi, \Phi (\frac{g_{SM}}{\sqrt{Y_2}},\frac{g_X}{\sqrt{Y_X}}, H, \varphi)  \chi \rangle 
\nonumber \\
&\stackrel{!}{=}& \langle \chi, \Phi (g_{SM},g_X, H, \varphi)  \chi \rangle. 
\ee
This implies the boundary conditions 
\bb
\boxed{
Y_2(\Lambda) = Y_X(\Lambda) = \frac{2 \pi^2}{f_4} = (4+ \tfrac23 N) \, g_2(\Lambda)^2 
}
\label{YukawasatLambda}
\ee
where the last equality follows from \eqref{SU2atLambda}. 

\paragraph{Scalar quartic coupling relations}

To obtain the  boundary conditions for the quartic couplings $\lambda_1$, $\lambda_2$ and $\lambda_3$ of the scalar fields we need
to compute the quartic part of $a_4(D_\Phi^2)$  for the mass dimension one scalar fields $\tilde H$ and $\tilde \varphi$. Inserting again
into the Spectral Action \eqref{SpectralAction} and equating with the standard normalisation yields the desired boundary conditions. To simplify
matters we split the computation into three parts for each of quartic couplings $\lambda_1$, $\lambda_2$ and $\lambda_3$.
\medskip

\noindent
Restricting the Spectral Action to the term proportional to $\tilde H^4$ we find  
\bb
\frac{f_4}{8 \pi^2} \int_M \tr_\HH(\Phi^4) |_{\tilde H^4} \,dx &=& \frac{4 f_4}{8 \pi^2}  \G_2 \int_M \tr[(\tilde H^*\tilde H)^2]  \,dx
\nonumber \\
&=& \frac{2 \pi^2}{f_4} \frac{ \G_2}{\Y_2^2} \int_M |H|^4 \,dx
\nonumber \\
&=& \frac{2 \pi^2}{f_4} \frac{ G_2}{Y_2^2} \int_M |H|^4 \,dx
\nonumber \\
&\stackrel{!}{=}& \frac{\lambda_1}{6} \int_M |H|^4 \,dx
\ee
where the traces of the fourth powers of the Yukawa mass matrices $\G_2$ and the Yukawa coupling matrices $G_2$ are defined in Appendix B, see 
\eqref{fourthpowermasssums} and \eqref{fourthpowersums}. The mass scale has been divided out in order to obtain the standard normalisation 
of the Yukawa  couplings \cite{MV83,MV84,MV85}.
The boundary condition of  $\lambda_1$ which is associated to the quartic potential of the scalar field $H$
\bb
\boxed{
\lambda_1(\Lambda) = 6 \frac{2 \pi^2}{f_4} \frac{ G_2(\Lambda)}{Y_2(\Lambda)^2} = g_2(\Lambda)^2 (24 + 4 N) \frac{ G_2(\Lambda)}{Y_2(\Lambda)^2}
}
\label{lambda1atLambda}
\ee
follows then with \eqref{SU2atLambda}. Restricting to the term proportional to $\tilde \varphi^4$ gives
\bb
\frac{f_4}{8 \pi^2} \int_M \tr_\HH(\Phi^4) |_{\tilde \varphi^4} \,dx  &=& \frac{2 \pi^2} {f_4} \frac{ G_X}{Y_X^2} \int_M | \varphi|^4 \,dx
\nonumber \\
&\stackrel{!}{=}& \frac{\lambda_2}{6} \int_M| \varphi|^4\,dx
\ee
which yields, again with \eqref{SU2atLambda} the boundary condition of $\lambda_2$:
\bb
\boxed{
\lambda_2(\Lambda) = 6 \frac{2 \pi^2}{f_4} \frac{ G_X(\Lambda)}{Y_X(\Lambda)^2} = g_2(\Lambda)^2 (24 + 4 N) \frac{ G_X(\Lambda)}{Y_X(\Lambda)^2}
}
\label{lambda2atLambda}
\ee
Restricting the Spectral Action to the term proportional to $\tilde H^2 \, \tilde \varphi^2$   
\bb
\frac{f_4}{8 \pi^2} \int_M \tr_\HH(\Phi^4) |_{\tilde H^2 \tilde \varphi^2} \,dx  &=& \frac{4 \pi^2}{f_4} \frac{ G_{\nu X_1}}{Y_2 Y_X} 
\int_M |H|^2 | \varphi|^2 \,dx
\nonumber \\
&\stackrel{!}{=}& \frac{\lambda_3}{3} \int_M  |H|^2 |\varphi|^2\,dx
\ee
we obtain the boundary condition for the coupling $\lambda_3$ of  
the interaction term for the two scalar fields:
\bb
\boxed{
\lambda_3(\Lambda) = 3 \frac{4 \pi^2}{f_4} \frac{ G_{\nu X_1}(\Lambda)}{Y_2(\Lambda)Y_X(\Lambda)} = g_2(\Lambda)^2 (24 + 4 N) 
\frac{ G_{\nu X_1}(\Lambda)}{Y_2(\Lambda)Y_X(\Lambda)} 
}
\label{lambda3atLambda}
\ee

\subsection{Dimensionful parameters} 

Let us now turn to the dimensionful parameters in  the Spectral Action. While the status of the dimensionless parameters in Euclidean
quantum field theories, e.g. their flow under the renormalisation group equations is fairly well understood, this is not at all the case for the dimensionful parameters
such as the negative scalar ``mass'' terms, the Planck mass and the cosmological constant. For these terms the Spectral Action also provides boundary values
at the cut-off scale $\Lambda$, but their physical significance is far from clear. Nevertheless it may be informative to calculate the specific values, especially the
Planck mass could give an idea whether the specific model aims into a reasonable direction.

\paragraph{Scalar field mass terms}

For the negative ``mass'' terms of the scalar fields with mass dimension one, we have to take the $a_2(D_\Phi^2)$ coefficient as well as the $a_4(D_\Phi^2)$ coefficient 
into account. Restricting the Spectral Action to terms proportional to $\tilde H^2$ only,  we find with \eqref{Yukawamassrelations} 
\bb
f_2 \Lambda^2 a_2(D_\Phi)|_{\tilde H^2} + f_4 a_4(D_\Phi)|_{\tilde H^2}  &=& - \frac{f_2}{4 \pi^2}  \Lambda^2 \int_M \tr_\HH (\Phi^2)|_{\tilde H^2} \, dx 
+ \frac{f_4}{8 \pi^2} \int_M \tr_\HH(\Phi^4)|_{\tilde H^2} \, dx
\nonumber \\
&=&  \left(  - \frac{f_2}{\pi^2}  \Lambda^2 \Y_2+  \frac{2\, f_4}{2 \pi^2}   \tr (\Gamma_\nu^2 M_\nu^* M_\nu) \right) \int_M \tr(\tilde H^* \tilde H) \, dx 
\nonumber \\
&=&  \left(  - \frac{2 \, f_2}{f_4}  \Lambda^2 +  \frac{  2  \tr (\Gamma_\nu^2 g_\nu^* g_\nu)}{Y_2} \right) \int_M |H|^2 \, dx 
\nonumber \\
&\stackrel{!}{=}&\int_M - \mu_H^2 |H|^2 \, dx,
\ee
and if we restrict to terms proportional to $\tilde \varphi^2$ we get

\bb
f_2 \Lambda^2 a_2(D_\Phi)|_{\tilde \varphi^2} + f_4 a_4(D_\Phi)|_{\tilde \varphi^2}  
&=& - \frac{f_2}{4 \pi^2}  \Lambda^2 \int_M \tr_\HH (\Phi^2)|_{\tilde \varphi^2} \, dx 
+ \frac{f_4}{8 \pi^2} \int_M \tr_\HH(\Phi^4)|_{\tilde \varphi^2} \, dx
\nonumber \\
&=&  \left(  - \frac{f_2}{\pi^2}  \Lambda^2 \Y_X+  \frac{2\, f_4}{2 \pi^2}   \tr (M_{X_1}^* M_{X_1} M_{\nu X_1}^* M_{\nu X_1}) 
\right) \int_M |\tilde \varphi|^2 \, dx 
\nonumber \\
&=&  \left(  - \frac{2 \, f_2}{f_4}  \Lambda^2 +  \frac{  2  \tr (M_{X_1}^* M_{X_1}  g_{\nu X_1}^* g_{\nu X_1})}{Y_X} \right) \int_M |\varphi|^2 \, dx 
\nonumber \\
&\stackrel{!}{=}&\int_M - \mu_\varphi^2 |\varphi|^2  \, dx.
\ee
One then reads off the following relations for $\mu_1$ and $\mu_2$ which are taken to be valid at the cut-off:
\bb
\mu_H^2 = \frac{2 \, f_2}{f_4}  \Lambda^2 - \frac{  2  \tr (\Gamma_\nu^2 g_\nu^* g_\nu)}{Y_2}, \qquad 
\mu_\varphi^2 = \frac{2 \, f_2}{f_4}  \Lambda^2 -  \frac{  2  \tr (M_{X_1}^* M_{X_1}  g_{\nu X_1}^* g_{\nu X_1})}{Y_X}.
\label{negativmass}
\ee
\paragraph{Planck mass}

To obtain a constraint for the Planck mass (or equivalently for the gravitational coupling) at the cut-off scale, we need to consider again the 
$a_2(D_\Phi^2)$-term and the $a_4(D_\Phi^2)$-term. Restricting the Spectral Action to the terms proportional to the scalar curvature $R^g$, we find
with the relations \eqref{Yukawamassrelations} that
\bb
 f_2 \Lambda^2 a_2(D_\Phi^2)|_{R^g}  + f_4 a_4(D_\Phi^2)|_{R^g \tilde H^0,R^g \tilde \varphi^0} 
&=& \left( - \frac{\rk(\HH) \, f_2}{48 \pi^2} \Lambda^2  + \frac{f_4}{48 \pi^2} \tr_\HH (\Phi^2)|_{\tilde H^0,\tilde \varphi^0} \right)
 \int_M R^g \, dx
\nonumber \\
&\stackrel{!}{=}& -\frac{m_P^2}{16 \pi} \int_M R^g \, dx.
\ee
From this we get
\bb
m_P^2 =  \frac{\rk(\HH) \, f_2}{3 \pi} \Lambda^2  - \frac{2\,  f_4}{3\, \pi} 
\left( 2 \tr (M_{X_1}^*M_{X_1}) + 2 \tr (M_{V_c}^*M_{V_c}) + 2  \tr (M_{V_w}^*M_{V_w})  +  \tr (\Gamma_\nu^2)    \right)
\label{Planckmass}
\ee
for the Planck mass $m_P(\Lambda)$. 

\paragraph{Cosmological constant}

To calculate the term  in the Spectral Action which can be considered to play the r\^ole of  the cosmological constant, we need also need to take 
the $a_0(D_\Phi^2)$ coefficient into account. Restricting the Spectral Action to the terms proportional to the volume of the manifold we impose
\bb
f_0 \Lambda^4 a_0(D_\Phi)|_{\tilde \varphi^2} + f_2 a_2(D_\Phi)|_{\tilde H^0, \tilde \varphi^0} + f_4 a_4(D_\Phi)|_{\tilde H^0, \tilde \varphi^0}  
\stackrel{!}{=} \frac{2 \, \Lambda_c m_P^2}{16 \pi} \int_M \, dx
\ee
which gives implicitly the value of the ``cosmological constant'' $\Lambda_c$ at the cut-off scale,
\bb
\frac{2 \, \Lambda_c m_P^2}{16 \pi}  &=&  \frac{\rk(\HH)\, f_0}{4 \pi^2} \Lambda^4   - \frac{f_2}{2 \pi^2} \Lambda^2 \left( 2 \tr (M_{X_1}^*M_{X_1}) + 2 \tr (M_{V_c}^*M_{V_c}) 
 + 2 \tr (M_{V_w}^*M_{V_w})  +  \tr (\Gamma_\nu^2) \right) 
\nonumber \\
&&   + \frac{f_4}{4 \pi^2} \left(  \tr (\Gamma_\nu^4) + 2  \tr [(M_{X_1}^*M_{X_1})^2]  + 2  \tr [(M_{V_c}^*M_{V_c})^2]  + 2  \tr [(M_{V_w}^*M_{V_w})^2]  
\right).
\ee
We did not take into account terms coming from the potential of the scalar fields that would be induced by the symmetry braking. This has been done for
example in \cite{JKSS07}. 

\section{A numerical example \label{NE}}

To get an idea of the phenomenological consequences of the present model, we will pick a ``convenient'' point in the parameter space
of the model. The parameter space is spanned, in addition to the usual Standard Model parameters, by 
\begin{itemize}
\item the number of $X$- and $V$-generations $N$,
\item the $U(1)_X$-gauge coupling $g_4$,
\item the Neutrino-X-sector Yukawa coupling matrix $g_{\nu X_1}$,
\item the X-particle Yukawa matrices $\M_{X_1}$, $g_{X_2}$ and $g_{X_3}$,
\item the V-particle Dirac mass matrices $M_{V_c}$ and $M_{V_w}$
\end{itemize}
We assume that we have $N=3$ generations for aesthetic reasons (i.e. since there are three known Standard Model generations), so $\rk(\HH)=192$.
This fixes the conditions of the gauge couplings at the cut-off scale $\Lambda$ in \eqref{AllUatLambda} to   $\tfrac76 \; g_1^2 = g_2^2=g_3^2 = \tfrac43 \; g_4^2$. 
Furthermore we assume that the Standard Model Yukawa couplings are dominated by the top-quark coupling $y_t$ and the $\tau$-neutrino coupling $y_{\nu_\tau}$.
The Yukawa couplings involving the $X$-particles are dominated by the coupling of the $\tau$-neutrino to one generation of the $X_1$-particles $y_{\nu_\tau X_1^1}$.  
For the squared sums of the Yukawa matrix traces \eqref{squaredsums} and the sums of the fourth powers of the traces \eqref{fourthpowersums} we get the simplifications 
\bb
Y_2 &\approx& 3 \, y_t ^2+ y_{\nu_\tau}^2, \qquad Y_X \approx y_{\nu_\tau X_1^1}^2, 
\nonumber \\
G_2 &\approx& 3 \, y_t ^4+ y_{\nu_\tau}^4, \qquad G_X \approx y_{\nu_\tau X_1^1}^4,  \qquad G_{\nu X_1} \approx y_{\nu_\tau}^2 y_{\nu_\tau X_1^1}^2. 
\eee
The Majorana masses of the neutrinos and the Dirac masses of the $X_1$-particles we put to $m_{X_1} \sim 10^{14}$GeV. This choice of the Majorana masses 
is reasonable to produce the correct effective neutrino masses via the seesaw-mechanism \cite{CCM07,JKSS07,CC10}.  For the masses of the three $V_c$-particles
we choose $m_{V_c} \sim 5.5 \times 10^{15}$GeV. The particular value of the $V_c$-particle's masses is ``tuned'' in order to allow for the boundary conditions \eqref{AllUatLambda}
to be met at $\Lambda$. Since the $V_c$-particles have colour- and hypercharge they will modify the running of the respective coupling constants above $5.5 \times 
10^{15}$GeV by the necessary amount. We put the masses of the $V_w$-particles to $\sim \Lambda$ so they do not contribute to the running of the couplings.
\\ 
\noindent
Defining $x := g_{\nu_\tau}/g_t$ the boundary relations \eqref{AllUatLambda},  \eqref{YukawasatLambda},
\eqref{lambda1atLambda}, \eqref{lambda2atLambda} and \eqref{lambda3atLambda} among the dimensionless parameters 
at the cut-off scale $\Lambda$ simplify to
\bb
\frac76 \; g_1^2 = g_2^2=g_3^2 = \frac43 \; g_4^2 =  \frac16 \, g_t^2 (3 + x^2)   = \frac16 \, y_{\nu_\tau X_1^1}^2
= \frac{1}{36} \frac{(3+ x^2)^2}{3+x^4} \, \lambda_1
= \frac{1}{36}\, \lambda_2
= \frac{1}{36} \frac{3+x^2}{x^2} \, \lambda_3
\label{simpleboundary}
\ee
So the remaining free parameter for this particular point in parameter space is the ratio $x$ of the $\tau$-neutrino Yukawa coupling $g_{\nu_\tau}$  and the
top-quark Yukawa coupling $g_t$. This parameter has to be chosen such that the low energy value of $g_t$ gives  the experimental value of the top-quark mass. All normalisations are chosen as in \cite{JKSS07}, i.e. the Standard Model fermion masses are given by $m_f = \sqrt{2} (g_f/g_2) m_{W^\pm}$
and the real scalar fields are normalised such that the relevant terms in the Lagrangian take the form  $(1/2) (\partial \varphi)^2 + (\lambda/24) \varphi^4$.

\paragraph{Renormalisation group equations}
In order to determine the cut-off scale $\Lambda$ and the low-energy values of the quartic couplings for the scalar fields we need the renormalisation 
group equations for all couplings in \eqref{simpleboundary}. Furthermore we need the renormalisation group equation for the Yukawa coupling of the 
$\tau$-neutrino. Following \cite{MV83,MV84,MV85} and \cite{FJSE93} we obtain the one-loop renormalisation group equations:
\bb
16 \pi^2 \, \beta_{\lambda_1} &=& \frac94 (g_1^4 + 2  \, g_1^2 g_2^2 + 3 \, g_2^4) - (3  \,g_1^2 + 9  \, g_2^2) \lambda_1 
+ 4 (3 \, y_t ^2+ y_{\nu_\tau}^2) \lambda_1
\nonumber \\
&&    -12 (3 \, y_t ^4+ y_{\nu_\tau}^4)  + 4  \, \lambda_1^2 + \frac23 \lambda_3^2
\\
16 \pi^2 \, \beta_{\lambda_2} &=& 36  \, g_4^4 - 12  \, g_4^2 \lambda_2 + 4  \, y_{\nu_\tau X_1^1}^2 \lambda_2 -12 \, y_{\nu_\tau X_1^1}^4 
+ \frac{10}{3} \lambda_2^2 + \frac43 \lambda_3^2
\\
16 \pi^2  \, \beta_{\lambda_3} &=&  - \frac12 (3  \, g_1^2 + 9   \, g_2^2+ 12  \, g_4^2) \lambda_3 
+ 2  \,(3 \, y_t ^2+ y_{\nu_\tau}^2 + y_{\nu_\tau X_1^1}^2) \lambda_3
- 9 \, y_{\nu_\tau}^2 y_{\nu_\tau X_1^1}^2
\nonumber \\
&&     + \frac43 \lambda_3^2 +2 \, \lambda_1 \lambda_3 +\frac43 \lambda_2 \lambda_3
\\
16 \pi^2 \, \beta_{y_t} &=& \left( - \sum_{i=1}^4 c^t_i \, g_i^2 + \frac52 (3 \, y_t^2 + y_{\nu_\tau}^2) \right) \, y_t, \quad c^t_i=\left[\frac{17}{12},\frac94,8,0\right]
\\
16 \pi^2 \, \beta_{y_{\nu_\tau}} &=& \left( - \sum_{i=1}^4 c^\nu_i \, g_i^2 + 3 \, y_t^2 + \frac52 y_{\nu_\tau}^2 + \frac12 y_{\nu_\tau X_1^1}^2 
\right) \, y_{\nu_\tau}, \quad c^t_i=\left[\frac34,\frac94,0,0\right]
\\
16 \pi^2 \, \beta_{y_{\nu_\tau X_1^1}} &=& \left( - \sum_{i=1}^4 c^\nu_i \, g_i^2 +  y_{\nu_\tau}^2 + 2 y_{\nu_\tau X_1^1}^2 
\right) \, y_{\nu_\tau X_1^1}, \quad c^t_i=\left[0,0,0,3\right]
\\
16 \pi^2 \, \beta_{g_i} &=& b_i g_i^3 \quad b_i = \left(\frac{41}{6} + \frac19 N_{V_c}, -\frac{19}{6}, -7 +  \frac23 N_{V_c},  \frac23 (2 \, N_{X_1} 
+ N_{X_2}+ N_{X_3})    \right)
\ee
where $N_{X_1},N_{X_2},N_{X_3},N_{V_c}$ denote the active number of the respective  particles. At each of the mass thresholds,
i.e. at the Majorana mass at $10^{14}$ GeV and the $V_c$ particle mass at $ 5.5 \times 10^{15}$ GeV, the respective 
particles are integrated out, according to the Applequist-Carrazzone decoupling theorem \cite{AC75}. 
In particular the running of the parameters below $10^{14}$ GeV is given by $ \beta_{\lambda_1}$,  $\beta_{\lambda_2}$,  $\beta_{\lambda_3}$,
$\beta_{y_t}$ and the gauge couplings $\beta_{g_i}$,  where only the active $X_2$-particles are taken into account and the inactive
Yukawa couplings $y_{\nu_\tau}$, $y_{\nu_\tau X_1^1}$ are put to zero.

\paragraph{Numerical results}

First we determine the cut-off scale $\Lambda$ with the renormalisation group equations for the gauge couplings $g_1$, $g_2$ and $g_3$. 
As experimental low-energy values we take \cite{PDG12}
\bb
g_1(m_Z)=0.3575, \quad g_2(m_Z)= 0.6514, \quad g_3(m_Z)= 1.221. 
\eee
The one-loop analysis shows that the high-energy boundary conditions for the gauge couplings,  \eqref{simpleboundary},
\bb
\frac76 \; g_1(\Lambda)^2 = g_2(\Lambda)^2=g_3(\Lambda)^2 = \frac43 \; g_4(\Lambda)^2
\eee
can be met for $\Lambda \approx 2.5 \times 10^{18}$ GeV where the last equality defines the value of $g_4$ at $\Lambda$.  The value of the cut-off scale 
is close to the reduced Planck mass  $m_P^{red.} := m_P/\sqrt{8 \pi} \approx 2.4 \times 10^{18}$ GeV. If we assume that we can also neglect the mass of the $V_w$-particles 
with respect to $\Lambda$ and we find with $f_2 \approx 1.2$ the correct value of Planck mass \eqref{Planckmass}:
\bb
m_P (\Lambda) \approx \sqrt{\frac{64 f_2}{\pi}} \; \Lambda \approx 1.2 \times 10^{19} \; {\rm GeV}.
\ee
Since $f_4 \approx 12$ it follows from \eqref{negativmass}  that the negative ``mass'' terms $\mu_1(\Lambda)$ and $\mu_2(\Lambda)$ are approximately  
one order of magnitude smaller than $m_P (\Lambda)$. Since one cannot be certain how these ``mass'' terms behave under renormalisation group
flow, we will consider them to be free parameters of the theory which have to be fixed by experiment, i.e. the $W^\pm$-boson mass and of the
Higgs mass. 
\medskip

\noindent
The running of the gauge couplings is shown in figure \ref{runninggraph1}. We note again that the mass of the $V_c$-particles has been chosen such
that the high-energy boundary conditions \eqref{simpleboundary} are satisfied for  a single value of $\Lambda$. There will probably be 
different mass patterns for the $V_c$- and the $V_w$-particles which allow to satisfy \eqref{simpleboundary} for all gauge couplings. 
It is certainly worthwhile to investigate this point further.
\begin{figure}
\begin{center}
\psfrag{t}[Bl][l][1][0]{ $t[$log(GeV)$]$}
\psfrag{gg1}[Bl][r][0.9][0]{ $\sqrt{7/6} g_1(t)$}
\psfrag{gg2}[Bl][l][0.9][0]{ $g_2(t)$}
\psfrag{gg3}[][l][0.9][0]{ $g_3(t)$}
\psfrag{gg4}[Bl][l][0.9][0]{ $\sqrt{4/3}g_4(t)$}
\includegraphics[scale=0.4]{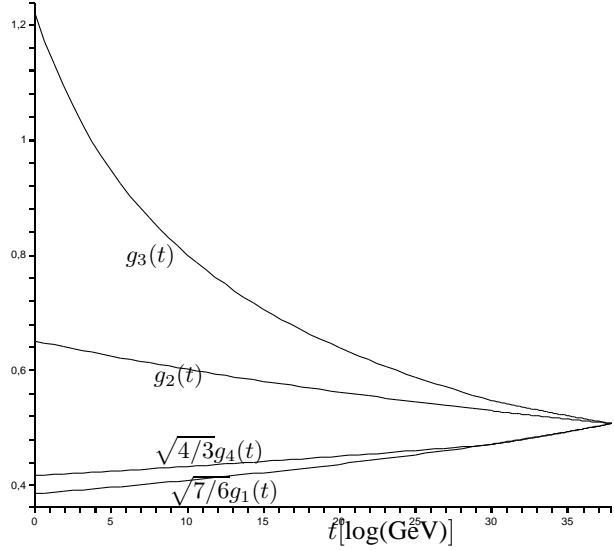}
\caption{Running of the gauge couplings with normalisation factors according to the high-energy boundary conditions \eqref{simpleboundary}.
The variable energy scale is defined by $E = m_Z \cdot \exp(t)$ with Z-Boson mass $m_Z = 91.19$ GeV.}
\label{runninggraph1}
\end{center}
\end{figure}
For the potential 
\bb
V(H,\varphi) =  -\mu_H^2 |H|^2 -\mu_\varphi^2 | \varphi|^2 +
\frac{\lambda_1}{6} \, |H|^4 +\frac{\lambda_2}{6} \, 
| \varphi|^4 +\frac{\lambda_3}{3} \, | H|^2 | \varphi|^2.
\label{pot}
\ee
and the present numerical values of the quartic couplings 
both scalar fields  have nonzero vacuum expectation values, $|\langle H \rangle| = v_H/\sqrt{2} \neq  0$ and
$|\langle \varphi \rangle| = v_\varphi/\sqrt{2} \neq 0$. 
The  vacuum expectation value $v_H$ of
the first scalar field can be determined by  the $W^\pm$ boson mass, 
$m_{W^{\pm}}=(g_2/2) \, v_H$.  With the experimental value $m_{W^{\pm}}=80.34$ GeV \cite{PDG12} we obtain 
$v_H = 246.8$ GeV.
In the numerical analysis it turns out that the errors are dominated by the experimental uncertainties of the top quark mass and
the presumable Higgs mass. Therefore we neglect the experimental uncertainties of the $W^\pm$ bosons and the gauge couplings
$g_1$, $g_2$ and $g_3$.
The vacuum expectation value of $\varphi$ is a free parameter and is
determined by $\mu_\varphi$.  We obtain the physical real scalars $h_H$ and $h_\varphi$ in the 
standard notation
\bb
H = \frac{1}{\sqrt{2}} \, \pp{0 \cr h_H + v_H}, \quad {\rm and}
\quad \varphi = \frac{1}{\sqrt{2}} \, (h_\varphi +v_\varphi)
\ee 
The mass matrix is not diagonal in the weak basis, but the
mass eigenvalues can be calculated to be \cite{EK07}
\bb
m_{\phi_H,\phi_\varphi} = \frac{\lambda_1}{6} \, v_H^2 + \frac{\lambda_2}{6} \, v_\varphi^2
\pm \sqrt{ \left(\frac{\lambda_1}{6} \, v_H^2 - \frac{\lambda_2}{6} \, v_\varphi^2 \right)^2
+ \frac{\lambda_3^2}{9} \, v_H^2 v_\varphi^2},
\label{masseigenvals}
\ee
where the real mass eigenstates $\phi_H$ and $\phi_\varphi$ are given by 
\bb
\pp{\phi_H \cr \phi_\varphi} = \pp{\cos \theta & - \sin \theta \cr \sin \theta & \cos \theta} 
\pp{h_H \cr h_\varphi}
\ee
and
\bb
\tan (2 \theta) = \frac{ 2 \lambda_3 v_H v_\varphi}{\lambda_1 v_H^2 - \lambda_2 v_\varphi^2}\, .
\ee
To obtain the low energy values of the quartic couplings $\lambda_1$, $\lambda_2$ and $\lambda_3$ we evolve their values at the cut-off $\Lambda$ fixed by the boundary 
conditions \eqref{simpleboundary}:
\bb
 g_2(\Lambda)^2 = \frac{1}{36} \frac{(3+ x^2)^2}{3+x^4} \, \lambda_1 (\Lambda)
= \frac{1}{36}\, \lambda_2 (\Lambda)
= \frac{1}{36} \frac{3+x^2}{x^2} \, \lambda_3 (\Lambda)
\eee
to low energies. In addition we have to take into account the running of the gauge couplings and the relevant Yukawa couplings. For the latter  \eqref{simpleboundary} 
implies at the cut-off
\bb
g_2(\Lambda)^2= \frac43 \; g_4^2 =  \frac16 \, g_t(\Lambda)^2 (3 + x^2)   = \frac16 \, y_{\nu_\tau X_1^1}(\Lambda)^2
\eee 
The top-quark mass has the experimental value $m_t =173.5 \pm 1.4$ GeV \cite{PDG12} which determines the paramter $x=2.145 \pm 0.065$.
As a very recent input we have the Higgs mass $m_{\Phi_H}=125.6 \pm 1.2$ GeV with the maximal combined experimental uncertainties of ATLAS \cite{ATLAS12} and 
CMS \cite{CMS12}. We identify the Higgs mass with the lower mass eigenvalue in \eqref{masseigenvals}. As it turns out, this is a reasonable 
assumption as long as the vacuum expectation value $v_\varphi$ is larger then the $v_H$ which is determined  by the $W^\pm$ boson mass
to be $v_H = 246.8$ GeV.
\medskip

\noindent
With the experimental values of the top quark and the $W^\pm$ boson masses and the previous results on the gauge couplings we obtain from 
the running of the couplings and the whole set of boundary conditions \eqref{simpleboundary} at $\Lambda$ from the mass eigenvalue equation
\eqref{masseigenvals} that $v_\varphi = 702 \pm 239$ GeV. The error in the vacuum expectation value is rather large due to the flat slope of the curve of the
light eigenvalue, see figure \ref{massEWs1}. The flat slope amplifies the experimental uncertainties of the top quark mass and
the Higgs mass, i.e. the lower vacuum expectation value. We notice that $v_\varphi$ is larger then $v_H$ but of the same order of magnitude. This 
justifies the assumption to run all three quartic couplings down to low energies and also ensures that the vacuum is stable up to the cut-off scale
$\Lambda \approx 2.5 \times 10^{18}$  GeV, see \cite{MEGLS12} and references therein.
For the second mass eigenvalue we find $m_{\Phi_\varphi} = 445 \pm 139$ GeV. For the mass of the new gauge boson $Z_X$ associated to the
broken $U(1)_X$ subgroup $m_{Z_X} =  254 \pm 87$ GeV where we took the value of $g_4(250 \; {\rm GeV})\approx 0.3617$ as a very good
approximation and neglected further uncertainties of $g_4$.
\medskip

\noindent
To get a rough estimate of the masses of the $X_2$- and $X_3$-particles we note that the $g_{\nu_\tau X_1^1} \sim 1$. So if we assume that
$g_{\nu_\tau X_1^1}$ dominates the sum of the squared Yukawa coupling $Y_X$ with an accuracy of a few percent, it is reasonable to take  $g_{X_{2/3}^i} \leq 0.1$ for the remaining Yukawa
couplings. This leads to $X_{2/3}$-particle masses of the order 50 GeV, or less. 

\begin{figure}
\begin{center}
\psfrag{AA}[Bl][l][1][0]{ $v_2$ $[{\rm GeV}]$}
\psfrag{BB}[][l][0.9][0]{ $m [{\rm GeV}]$}
\psfrag{mHH}[Bl][l][0.9][0]{ $m_H $}
\psfrag{mVV}[][l][0.9][0]{ $m_\varphi$}
\psfrag{mHreal}[Bl][l][0.9][0]{ $m_{exp.}$}
\includegraphics[scale=0.4]{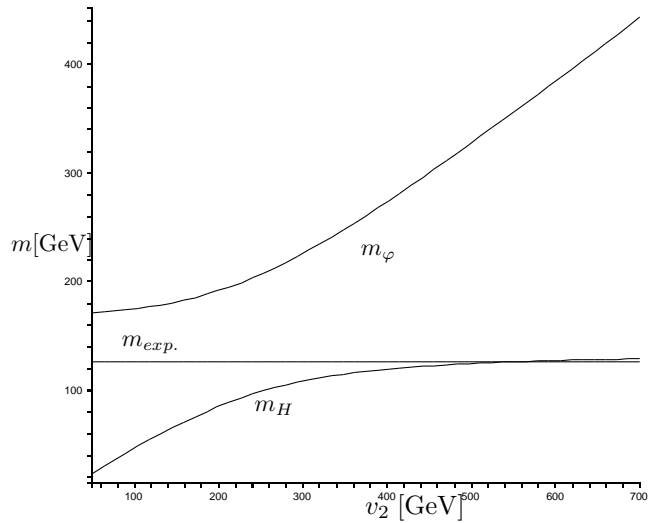}
\caption{Example for the mass eigenvalues $m_H$ and $m_\varphi$  of the scalar fields $H$ and $\varphi$ with respect
to $v_2$. Here the top quark mass is taken to be $m_t (m_Z) =173.2$ GeV and the experimental Higgs mass $m_{exp.}
= 126$ GeV. We obtain a heavy scalar with $m_\varphi = 347$ GeV and a $U(1)_X$-scalar boson mass 
$m_{Z_X} = 193$ GeV. }
\label{massEWs1}
\end{center}
\end{figure}

\section{Concluding remarks}

In the present paper we have investigated an extension of the almost-commutative model \cite{St09}. 
Let us summarise the main features of the present model are:
\begin{itemize}
\item it is fully compatible with the axioms of noncommutative geometry.
\item is is free of harmful Yang-Mills anomalies.
\item the high-energy boundary condition of the Spectral Action  for the Planck mass gives the correct
experimental value for a reasonable value of the second moment of the cut-off function.
\item the running of the dimensionless couplings is compatible with the high-energy boundary conditions
of the Spectral Action.
\item the model is compatible with the experimental value of the Higgs mass. It shares this property 
with the model in \cite{St09}.
\item the new scalar fields stabilises the scalar vacuum throughout the perturbative running of the couplings.
\item the model has a ``dark sector''. Its lighter particles are made out of bosons, $Z_X$ \& $\Phi_\phi$,
and fermions, $X_2$ \& $X_3$.  
\end{itemize}
Yet, we are left with many open questions. 
\begin{itemize}
\item How stable are the numerical results w.r.t. variations in the parameter space?
\item Are the new particles detectable, e.g. at the LHC?
\item Can particles from the ``dark sector'' of the model solve the dark matter problem?
\item The model is sensitive to the running of the top-quark mass. How do two-loop effects in 
the renormalisation group running effect the numerical results?
\item How compatible is the model with respect to high precision experimental data such as the anomalous magnetic
moment of the muon?
\item Does the model have a sizeable kinetic mixing of $U(1)_X$ to the hypercharge group $U(1)_Y$? 
\item How $Z'$-like is the $Z_X$-boson?
\end{itemize}
In view of the experimental value of the Higgs mass which has been discovered to be $\sim 125$ GeV \cite{ATLAS12,CMS12}
the key feature of this model is the introduction of  a new scalar field coupled to the Higgs scalar.
An extra scalar field \cite{CC12} can also be introduced via a ``grand symmetry'' breaking mechanism \cite{DLM13}. 
Here the fluctuations  generate a field replacing the Majorana mass for the right-handed neutrinos. The Yukawa coupling of at least one 
right-handed neutrino species the lepton doublet has to be $\sim 1$. By the same argument the Yukawa coupling 
of the neutrino to its anti-particle is also $\sim 1$ and it follows that for a cut-off energy $\Lambda \sim 10^{17}$ GeV 
the vacuum expectation value needs to be   $\sim 10^{14}$ GeV \cite{JKSS07}. Such a scenario makes the stabilisation 
of the scalar vacuum quite difficult \cite{MEGLS12} since the heavy scalar and the heavy neutrinos have to be integrated out at 
$\sim 10^{14}$ GeV to produce an effective potential. The situation may improve for lower cut-off energies and more 
complex patterns of the Yukawa matrices.
The model merits a closer analysis to see whether the running of the effective quartic coupling is not dominated too much
by the top quark Yukawa coupling, in order to produce an experimentally  acceptable value for the light scalar.
The more complicated Pati-Salam-model \cite{CCS13} with several additional scalar fields may have interesting phenomenological 
consequences. Here a detailed analysis will certainly prove to be most interesting. 
\medskip

\noindent
It is possible, yet far from trivial, to construct variations of the present model that may be similarly consistent 
(or inconsistent) with experimental data. In figure \ref{smallmodel} a reduced model is depicted containing only
two species of $X$-particles. A consequence is a smaller Hilbert space and stronger constraints on the representation
of the algebra. Demanding anomaly cancelation implies then that the $V_{c/w}$-particles acquire also $U(1)_X$ charge.
Therefore one would expect more kinetic mixing. This model will be investigated further in an upcoming publication \cite{St13}.
One could also imagine to extend the Standard Model group by a non-abelian group in the spirit of \cite{St07} and
\cite{SS07}. Such a model could also be constructed as a variation of the present model. In the Krajewski diagram 
this would amount to replacing the $X_{2/3}$-arrows by arrows in place of the dashed lines in figure \ref{Krajewski1}.
The phenomenological effects of such a model are certainly worth analysing.

\paragraph{Acknoledgements:} The author appreciates financial support from the SFB 647: {\it Raum-Zeit-Materie}
funded by the Deutsche Forschungsgemeinschaft.

\appendix

\section{Physical normalisation of the gauge fields}

We assume that $\HH$ is an associated vector bundle with  structure group
$G:= U(1)_1 \times \cdots \times U(1)_g \times SU(n_1) \times \cdots \times SU(n_h)$, $n_j \geq 2$. 
Furthermore we assume that $\HH$ splits as
\bb
\HH = \bigoplus_{k=1}^m \HH^k \qquad \HH^k = \bigotimes_{s=1}^{d_k} \HH^k_s
\ee
where the subspaces $\HH^k$ represent the particle multiplets and the subspaces $\HH^k_s$  carry trivial or 
fundamental sub-representations $\rho^s$ of the $U(1)_j$ or $SU(n_j)$ subgroups of  $G$.
\\ \\
\noindent
The covariant derivative restricted to $\HH^k_s$ is  $\nabla^\HH|_{\HH^k_s}\chi^k_s := \nabla^s_{e_i} \chi^k_s = \partial_{e_i} \chi^k_s + \rho^s(\omega(e_i)) \chi^k_s$ and  for a multiplet $\chi^k=\chi^k_1\otimes \cdots \otimes \chi^k_d \in \HH^k$ we have
\bb
\nabla^\HH |_{\HH^k}\chi^k &=& (\nabla^1 \chi^k_1)\otimes \chi^k_2 \otimes \cdots \otimes \chi^k_{d_k} 
 + \chi^k_1\otimes (\nabla^2 \chi^k_2) \otimes \cdots \otimes \chi^k_{d_k} 
\nonumber \\
&& + \cdots +  \chi^k_1\otimes \chi^k_2 \otimes \cdots \otimes (\nabla^{d_k} \chi^k_{d_k} )
\ee
For the curvature 2-form $\Omega^{\HH^k}:=\Omega^\HH|_{\HH^k} $ one has with the obvious notation
\bb
\Omega^{\HH^k} =  \Omega^1 \otimes  \cdots \otimes \id_{d_k} + \cdots +  \id_1 \otimes \cdots \otimes \Omega^{d_k} 
\ee
and since $\tr_{\HH^k_s} \Omega^s = 0$ 
\bb
\tr_{\HH^k} (\Omega^{\HH^k}\Omega^{\HH^k}) = \tr_{\HH^k} (\Omega^1 \Omega^1)  \tr_{\HH^k}(\id_2) \cdots \tr_{\HH^k}(\id_{d_k})
+ \cdots + \tr_{\HH^k}(\id_1) \cdots \tr_{\HH^k}(\id_{d_k-1})  \tr_{\HH^k} (\Omega^k \Omega^k) 
\ee
We now pass to the conventions that are usually used in the physics literature. For the $su(n_s)$ sub-Lie algebras
of the Lie algebra of $G$ we choose a basis $t_a$ with normalisation $\tr_{\HH} (\rho^s(t_a) \rho^s(t_b)) = -\tfrac12 \delta_{ab}$. In this basis we can write 
\bb
\tr_{\HH^k} (\Omega^s \Omega^s) = \sum_{a,b,i,j} (A^s)^a_{ij} (A^s)^b_{ij} \tr_{\HH^k} (\rho^s(t_a) \rho^s(t_b))
= - \frac12 \sum_{a,i,j} (A^s)^a_{ij} (A^s)^a_{ij}
\ee
For the $u(1)_s$ sub-Lie algebras case we normalise the basis  such that
\bb
\tr_{\HH^k} (\Omega^s \Omega^s) =   (q^s)^2 \sum_{i,j} (\Omega^s)_{ij} (\Omega^s)_{ij} = -  (q^s)^2 \sum_{i,j} (A^s)_{ij} (A^s)_{ij}  
\ee
where $q^s \in \Z$ is the charge of the representation $\rho^s$. 
For the Standard Model with structure group $G_{SM} = U(1)_Y \times SU(2)_w \times SU(3)_c$ it is customary to denote the
components of the curvature 2-forms with respect to these basis by $(A^Y)_{ij} =: B_{ij}$, $(A^w)^a_{ij} =: W^a_{ij}$
and $(A^c)^a_{ij} =: G^a_{ij}$ and to normalise the $U(1)_Y$-hypercharge of the right-handed electron to $q^{e_r} = -1$.

\section{Krajewski diagrams, representations \& fluctuations}

A real, finite spectral triple of $KO$-dimension six \cite{C94,C06,B07} is given by
($\mathcal{A},\HH,D, $ $J,\chi$)  with a finite dimensional real $*$-algebra
$\mathcal{A} $, a faithful representation
$\rho$ of $\mathcal{A}$ on  a finite dimensional complex Hilbert space $\HH$.
The Dirac operator $D$ is selfadjoint  and has a compact resolvent.Three
additional operators are defined on
$\HH$: the real structure $J$ is anti-unitary,  and the chirality
$\chi$ which is a unitary involution. These operators satisfy:

\begin{itemize}
\item $J^2=1$,
$[J,D]=\{ J,\chi\}=0$, $D \chi =-\chi D$ and $[\chi,\rho(a)]=0$ 
\item the zero order axiom, $[\rho(a),J\rho(b)J^{-1}]=0$,  $\forall a,b \in \mathcal{A}$.
\item the first order axiom, $[[D,\rho(a)],J\rho(b)J^{-1}]=0$,  $\forall a,b \in \mathcal{A}$
\item the orientability axiom which states that the chirality 
can be written as a finite sum $\chi =\sum_i\rho(a_i)J\rho(b_i)J^{-1}$.
\item  the Poincar\'e duality axiom which states that intersection form
$\cap_{ij}:=\tr_\HH (\chi \,\rho (p_i) J \rho (p_j) J^{-1})$ is non-degenerate,
$\rm{det}\,\cap\not=0$. The
$p_i$ are minimal rank projections in $\mathcal{A}$. 
\item the dimension axiom, the regularity axiom and the finiteness axioms, for details see e.g.
\cite{C94,C96,CM08}
\end{itemize}

\noindent
The internal or finite algebra is a finite sum of simple real, complex or quaternionic matrix
algebras.

\paragraph{The spectral triple}

\begin{figure}
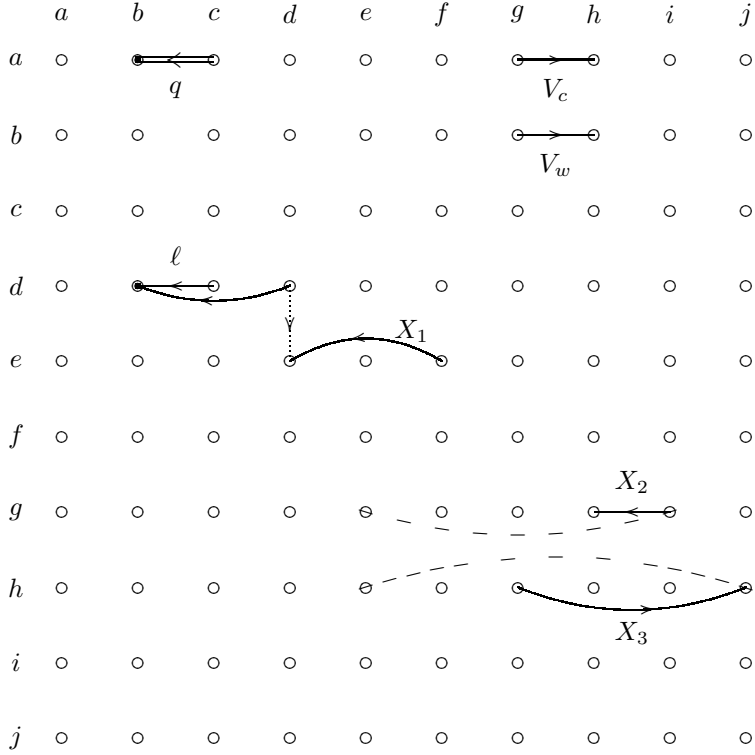

\begin{center}
\begin{tabular}{c}
\rxyzehn{0.4}{
,(10,-20);(15,-20)**\dir{-}?(.4)*\dir{<}
,(10,-20);(20,-20)**\crv{(15,-22)}?(.4)*\dir{<}
,(10,-20)*\cir(0.2,0){}*\frm{*}
,(12.5,-18)*{\ell}
,(10,-5);(15,-5)**\dir2{-}?(.4)*\dir2{<}
,(10,-5)*\cir(0.2,0){}*\frm{*}
,(12.5,-7)*{q}
,(20,-20);(20,-25)**\dir{..}?(.6)*\dir{>}
,(20,-25);(30,-25)**\crv{(25,-22)}?(.4)*\dir{<}
,(28,-23)*{X_1}
,(35,-5);(40,-5)**\dir{-}?(.6)*\dir{>}
,(37.5,-7)*{V_c}
,(35,-10);(40,-10)**\dir{-}?(.6)*\dir{>}
,(37.5,-12)*{V_w}
,(45,-35);(40,-35)**\dir{-}?(.6)*\dir{>}
,(42.5,-33)*{X_2}
,(35,-40);(50,-40)**\crv{(42.5,-43)}?(.6)*\dir{>}
,(42.5,-43)*{X_3}
,(45,-35);(25,-35)**\crv{~**\dir{--},(35,-38)}
,(25,-40);(50,-40)**\crv{~**\dir{--},(37.5,-36)}
}
\end{tabular}
\caption{Krajewski diagram of the extended Standard Model. The quarks are represented by the double arrow labelled $q$, the 
leptons by the split-double arrow $\ell$ and the $X$- and $V$-particles by the single  arrows with the respective label.
The dotted  line indicates the interaction term between the right-handed neutrino and
the left-handed $X_1$-particle induced by the new scalar field $\varphi$. The dashed lines denote the position of arrow for possible 
extensions by non-abelian gauge groups.} 
\label{Krajewski1} 
\end{center}
\end{figure}

The model under consideration in this paper can be realised as a spectral triple of KO-dimesion six \cite{C06,B07} and is constructed
from building blocks from the classification found in \cite{ISS04,JS05,Sc05b,JSS05,JS08,JS09} which is an specialisation of the general 
classification \cite{PS98,K98} to irreducible spectral triples. A predecessor of the model has been analysed in \cite{St09}. 
By its very construction from a Krajewski diagram
the model fulfils all mathematical axioms of a spectral triple, save the axiom of orientability which is violated by the right-handed neutrinos \cite{St06b}.
\medskip

\noindent 
We focus here on the finite part of the spectral triple since it encodes the particle content of the model as well as its dynamics. The 
matrix algebra of the spectral triple is  $ \mathcal{A} = M_3 (\C) \oplus M_2(\C) \oplus \C^8 \ni (a,b,c,d,e,f,g,h,i,j)$ and its representation
can be read off the Krajewski diagram in figure \ref{Krajewski1}. The model is constructed from diagram 3 in \cite{JS09} and diagram 5 in \cite{JS08}. 
These  diagrams cover for the Standard Model particles, save the right-handed neutrinos, and the $X$-particles. The $V_c$- and the $V_w$- particles
have been added ``by hand'' and are just as the right-handed neutrinos ``reducible'' in the sense of \cite{ISS04}.
\medskip

\noindent
To facilitate the deciphering we split the representation into the Standard Model
sub-representation, the sub-representation of the $X$-particles and the sub-representation of the $V_c$- and the $V_w$- particles:
\bb
\rho = \rho_{SM} \oplus \rho_X \oplus \rho_{V_{c/w}} \oplus \overline{\rho^c_{SM}} \oplus \overline{ \rho^c_X} \oplus \overline{ \rho^c_{V_{c/w}}}. 
\ee
Furthermore we split into left- and right-handed sub-representations, $\rho = \rho_\ell \oplus \rho_r$, and find:
\bb
&&\rho_{SM,\ell} = \diag [b \otimes 1_3, b], \quad \rho_{SM,r} = \diag[c \otimes 1_3, \bar c \otimes 1_3 , \bar c, \bar d],  
\nonumber \\
&&\rho^c_{SM,\ell} = \diag[1_2 \otimes a, d \, 1_2], \quad \rho^c_{SM,r} = \diag [a,a,d,d],
\nonumber \\
&&\rho_{X,\ell} = \diag[\bar d , i ,  g ], \quad \rho_{X,r} = \diag [f , h, j ], 
\nonumber \\ 
&& \rho^c_{X,\ell} = \diag [e, \bar g , \bar h], \quad \rho^c_{X,r} = \diag[ e, \bar g, \bar h], 
\nonumber \\
&& \rho_{V_{c/w},\ell} = \diag [  g \otimes 1_3, g \otimes 1_2], \quad  \rho_{V_{c/w},\ell} = \diag [ h \otimes 1_3, h \otimes 1_2], 
\nonumber \\
&& \rho^c_{V_{c/w},\ell} = \diag[ \bar a , \bar b], \quad  \rho^c_{V_{c/w},\ell} = \diag[ \bar a, \bar b],    
\ee
From this representation we can construct a lift and a central extension of the non-abelian part of the unitary group 
$\mathcal{U}^{na} = U(2) \times U(3)$.  Denoting elements in the two subgroups by $v\in U(2)$ and $u \in U(3)$ we
identify the determinant of the $U(2)$ subgroup with the hypercharge subgroup $ U(1)_Y \ni \det(v)$.
Furthermore we have a new subgroup   $ U(1)_X \ni \det(u)$ from the determinant of $U(3)$. 
The elements of the special unitary groups are then $\tilde v \in SU(2)_w$ and $\tilde u \in SU(3)_c$ where we identified these
subgroups with the weak- and the colour-subgroup.

\paragraph{Unitary lift and fluctuations}
The reduction of the unitary group of the algebra to the Standard Model group with extra $U(1)_X$,  
$G= U(1)_Y \times SU(2)_w \times SU(3)_c \times U(1)_X$ is obtained by lifting $\mathcal{U}^{na} = U(2) \times U(3)$
to the Hilbert space  and centrally extending \cite{LS01} by $U(1)_Y$ and $U(1)_X$. 
Note that we changed the order of the subgroups in $G$ with respect to the order of the subalgebras in $\mathcal{A}$
to be in accord with the standard enumeration of the gauge couplings $g_1$, $g_2$ and $g_3$.
The central extension is chosen such that the particle model is 
free of harmful Yang-Mills anomalies. For the present model the lift $L = L_\ell \oplus L_r \oplus \overline{L_\ell} \oplus \overline{L_r}$
can again be decomposed into the sub-representations of the Standard Model, the $X$- and the $V$-particles:
\bb
L(\det (u), \det(v) , u, v) = L_{SM} (\det (v) , \tilde u , \tilde  v)  \oplus L_X ( \det (u)) \oplus L_{V_{c/w}} (\det (v)  ,\tilde u, \tilde v)
\ee
Then the details of the lift for representations  on the sub-Hilbert spaces  are
\bb
&&L_{SM,\ell} (\det (v) , \tilde u , \tilde  v) = \diag[  \det(v)^{+1/6} \, \tilde v \otimes \tilde u, \det(v)^{-1/2} \, \tilde v],
\nonumber \\
&& L_{SM,r}   (\det (v) , \tilde u ) = \diag [\det(v)^{+2/3}\,  \tilde u, \det(v)^{-1/3} \, \tilde u, \det(v)^{-1}, \det(v)^0 ],
\nonumber \\
&&L_{X,\ell} (\det(u)) = \diag [\det(u)^{+1}, \det(u)^{+1}, \det(u)^{+1}],   
\nonumber \\
&& L_{X,r} (\det(u)) = \diag [\det(u)^{+1}, \det(u)^0 ,\det(u)^0]   
\nonumber \\
&& L_{V_{c/w},\ell} (\det (v) ,\det(u) ,\tilde u, \tilde v) = L_{V_{c/w},r} (\det (v) ,\det(u) ,\tilde u, \tilde v) = \diag [\det(v)^{-1/6} \, \overline{\tilde u},  \overline{\tilde v}]
\ee
where the central charges for the central extensions are given in the representations of the fermions in \eqref{repsfermions}. This representation
of the structure group $G$ is anomaly free due to the left-right symmetry in the $X$- and $V$-particle sector and the fact that the Standard Model
is anomaly free.
Writing $L_i :=  L(\det (u_i), \det(v_i) , u_i, v_i)$  and normalising  $\sum_i \, r_i =1$ we have  the following fluctuations of the Dirac operator
\bb
D_\HH &=&  \sum_i \,  r_i \, (\id_{\Sigma M} \otimes  L_i ) \, \slashed{\partial} \, ( \id_{\Sigma M} \otimes L_i^{-1}) 
\nonumber \\
&=& \sum_i \, r_i (\id_{\Sigma M} \otimes L_i) \, \sum_{j=1}^4\left(e_j\cdot \nabla^g_{e_j} \otimes \id_\HH 
\right) \, (\id_{\Sigma M} \otimes L_i^{-1} )
\nonumber \\
&&+  \sum_i \, r_i (\id_{\Sigma M} \otimes L_i) \, \sum_{j=1}^4\left(e_j\cdot  \otimes  \, \partial_{e_j}
\right) \, (\id_{\Sigma M} \otimes L_i^{-1})  
\nonumber \\
&=& \sum_{j=1}^4 \Big( e_j\cdot \nabla^g_{e_j} \otimes \id_\HH + \,
e_j \cdot \; \otimes \; \big( \partial_{e_i } + \sum_i \, r_i \,  L_i (\partial_{e_j} L_i ^{-1})  \big) \Big)
\nonumber \\
&=&\sum_{j=1}^4\left(e_j\cdot \nabla^g_{e_j} \otimes \id_\HH +
e_j \cdot \; \otimes \nabla^\HH_{e_j}  \right) 
\ee
From the fluctuations of the mass matrix 
\bb
\M = \pp{\Delta & \Gamma \\ \Gamma & \bar \Delta}
\ee
we get the Higgs endomorphism  
\bb
\Phi = \sum_i \,  r_i \, L(\det (u_i), \det(v_i) , u_i, v_i) \, \M \, L(\det (u_i), \det(v_i) , u, v_i)^{-1}.  
\ee
The sub-matrices of $\M$ are $\Gamma= 0 \oplus \Gamma_\nu \oplus 0$, the Majorana mass matrix for the right-handed neutrinos, and
$\Delta = \Delta_{SM,X_1}  \oplus \Delta_{X_{2/3},V}$, where the details can again be read off from the Krajewski diagram in figure 
\ref{Krajewski1}:
\bb
\Delta_{SM,X_1} = \pp{M_u \otimes 1_3 & M_d \otimes 1_3 & 0 & 0 & 0   \\ 0 & 0 & M_e & M_\nu & 0 \\
0 & 0 & 0 & M_{\nu X_1} & M_{X_1}  }, 
\quad \Delta_{X_{2/3},V} =  \pp{M_{X_1} & 0 & 0 & 0 \\ 0 & M_{X_2} & 0 & 0 \\ 0 & 0 & M_{V_c} & 0 \\ 0 & 0 & 0 & M_{V_w}},
\ee
and $M_u, M_d, M_e, M_\nu \in M_{6\times 3}(\C)$,  $M_{\nu X_1}, M_{X_{1/2/3}},M_{V_{c/w}} \in M_{3\times 3}(\C)$. 
The Higgs endomorphism after fluctuation is 
\bb
\Phi = \pp{\tilde \Delta & \Gamma \\ \Gamma & \overline{(\tilde \Delta)}},
\ee
where the details of $\tilde \Delta =\tilde  \Delta_{SM,X_1}  \oplus \tilde \Delta_{X_{2/3},V}$ in terms of the mass dimension zero scalar 
fields $\tilde H$ and $\tilde \varphi$ as well as the Yukawa mass matrices are as follows:
\bb
\tilde \Delta_{SM,X_1} &=&  \pp{\tilde H M_u \otimes 1_3 & \tilde H M_d \otimes 1_3 & 0 & 0 & 0   \\ 0 & 0 & \tilde H M_e & \tilde H M_{\nu} & 0 \\
0 & 0 & 0 & \tilde \varphi M_{\nu X_1} & M_{X_1}  },
\nonumber \\
\tilde \Delta_{X_{2/3},V_{c/w}} &=&  \pp{ \tilde \varphi  M_{X_2} & 0 & 0 & 0 \\ 0 & \tilde \varphi  M_{X_3} & 0 & 0 \\ 0 & 0 & M_{V_c} & 0 \\ 0 & 0 & 0 & M_{V_w}}.
\ee
One notices the appearance of the scalar field $\tilde \varphi$ in addition to the usual Higgs field $\tilde H$. Of special importance is the term
$ \tilde \varphi M_{\nu X_1}$ in  $\tilde \Delta_{SM,X_1} $ because it will generate with $\tilde H M_{\nu}$ the quartic $\tilde \varphi$-$\tilde H$
coupling in the spectral action. Therefore it is the main source to obtain a correct value for the Higgs mass \cite{St09}.

\paragraph{Traces of the Higgs endomorphism}
To calculate the Spectral Action \eqref{SpectralAction} we need the traces of the squares and the fourth powers of the Higgs endomorphism.
For the trace of the square of $\Phi$ we find
\bb
\tr_\HH (\Phi^2) = 4\, \Y_2 \tr(\tilde H^*\tilde H)+ 4 \, \Y_X |\tilde \varphi|^2+ 4 \tr (M_{X_1}^*M_{X_1}) + 4 \tr (M_{V_c}^*M_{V_c})
+ 4  \tr (M_{V_w}^*M_{V_w})  + 2 \tr (\Gamma_\nu^2) 
\ee
with the convenient abrevations
\bb
\Y_2 &:=& 3 \tr(M_u^* M_u) + 3 \tr(M_d^* M_d) + \tr(M_e^* M_e)  + \tr(M_{\nu}^* M_{\nu})
\nonumber \\
\Y_X&:=&  \tr(M_{\nu X_1}^* M_{\nu X_1})+\tr( M_{X_2}^* M_{X_2})  +\tr( M_{X_3}^* M_{X_3}).
\label{squaredmasssums}
\ee
Dividing $\Y_2$ and $\Y_X$ by a generic mass scale which is given by the normalisation of the 
Yukawa couplings  we find the standard traces of the squares of the Yukawa matrices:
\bb
Y_2 &:=& 3 \tr(g_u^* g_u) + 3 \tr(g_d^* g_d) + \tr(g_e^* g_e)  + \tr(g_{\nu}^* g_{\nu})
\nonumber \\
Y_X&:=&  \tr(g_{\nu X_1}^* g_{\nu X_1})+\tr( g_{X_2}^* g_{X_2}) +\tr( g_{X_3}^* g_{X_3}). 
\label{squaredsums} 
\ee
To calculate the trace of the fourth power of $\Phi$ we take into account that  $ \tr (\Gamma_\nu M_\nu^* \Gamma_\nu M_\nu) =0$ and get
\bb
\tr_\HH (\Phi^4) &=& 4 \, \G_2  \tr[(\tilde H^*\tilde H)^2]  + 4\, \G_X |\tilde \varphi|^4 + 8 \, \G_{\nu X_1} \tr(\tilde H^*\tilde H) |\tilde \varphi|^2
\nonumber \\
&& + 8  \tr (\Gamma_\nu^2 M_\nu^* M_\nu)\tr(\tilde H^*\tilde H) + 8 \tr  (M_{X_1}^* M_{X_1} M_{\nu X_1}^* M_{\nu X_1})  |\tilde \varphi|^2
\nonumber \\
&& + 2 \tr (\Gamma_\nu^4) + 4  \tr [(M_{X_1}^*M_{X_1})^2]  + 4  \tr [(M_{V_c}^*M_{V_c})^2]  + 4  \tr [(M_{V_w}^*M_{V_w})^2]  
\nonumber
\ee
again introducing convenient abbreviations
\bb
\G_2 &:=& 3 \tr[(M_u^* M_u)^2] + 3 \tr[(M_d^* M_d)^2] + \tr[(M_e^* M_e)^2]  + \tr[(M_{\nu}^* M_{\nu})^2]
\nonumber \\
\G_X&:=&  \tr[(M_{\nu X_1}^* M_{\nu X_1})^2]+\tr[( M_{X_2}^* M_{X_2})^2]  +\tr[( M_{X_3}^* M_{X_3})^2]
\nonumber \\
\G_{\nu X_1} &:=& \tr (M_\nu^* M_\nu M_{\nu X_1}^* M_{\nu X_1}) 
\label{fourthpowermasssums}
\ee
which are connected to the standard traces of the fourth powers of the Yukawa matrices if the normalisation mass
scale is divided out:
\bb
G_2 &:=& 3 \tr[(g_u^* g_u)^2] + 3 \tr[(g_d^* g_d)^2] + \tr[(g_e^* g_e)^2]  + \tr[(g_{\nu}^* g_{\nu})^2]
\nonumber \\
G_X&:=&  \tr[(g_{\nu X_1}^* g_{\nu X_1})^2]+\tr[( g_{X_2}^* g_{X_2})^2]  +\tr[( g_{X_3}^* g_{X_3})^2]  
\nonumber \\
G_{\nu X_1} &:=& \tr (g_\nu^* g_\nu g_{\nu X_1}^* g_{\nu X_1}). 
\label{fourthpowersums}
\ee

\begin{figure}
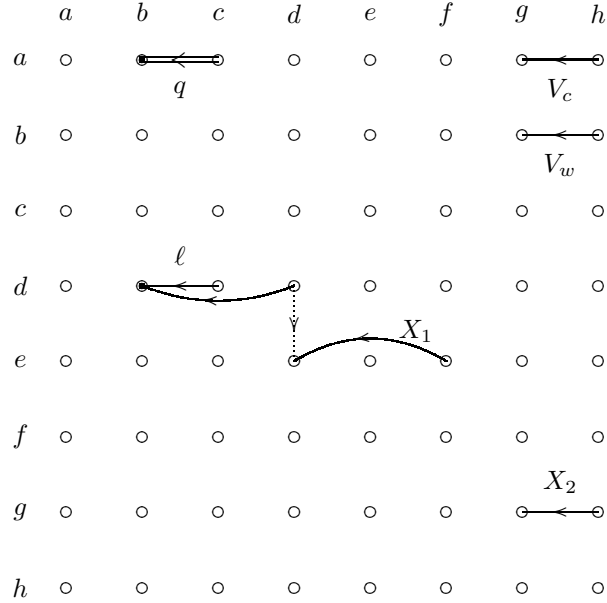

\begin{center}
\begin{tabular}{c}
\rxyacht{0.4}{
,(10,-20);(15,-20)**\dir{-}?(.4)*\dir{<}
,(10,-20);(20,-20)**\crv{(15,-22)}?(.4)*\dir{<}
,(10,-20)*\cir(0.2,0){}*\frm{*}
,(12.5,-18)*{\ell}
,(10,-5);(15,-5)**\dir2{-}?(.4)*\dir2{<}
,(10,-5)*\cir(0.2,0){}*\frm{*}
,(12.5,-7)*{q}
,(20,-20);(20,-25)**\dir{..}?(.6)*\dir{>}
,(20,-25);(30,-25)**\crv{(25,-22)}?(.4)*\dir{<}
,(28,-23)*{X_1}
,(35,-5);(40,-5)**\dir{-}?(.4)*\dir{<}
,(37.5,-7)*{V_c}
,(35,-10);(40,-10)**\dir{-}?(.4)*\dir{<}
,(37.5,-12)*{V_w}
,(35,-35);(40,-35)**\dir{-}?(.4)*\dir{<}
,(37.5,-33)*{X_2}
}
\end{tabular}
\caption{Krajewski diagram of the extended Standard Model. The
dotted  line indicates  the Dirac mass term leading to the new
scalar field $\varphi$.}
\label{smallmodel}
\end{center}
\end{figure}

%
%

\end{document}